\documentclass[11pt]{article}

\usepackage[T1]{fontenc}
\usepackage[cp1251]{inputenc}
\usepackage{textcomp}
\usepackage[centertags]{amsmath}
\usepackage{amsfonts}
\usepackage{amssymb}
\usepackage[pdftex]{hyperref}
\usepackage{graphicx}
\usepackage{graphbox}
\usepackage[numbers,sort&compress]{natbib}

\usepackage{paperinitial}

\paperinitialization{15mm}{15mm}{15mm}{15mm}{2pt}{10pt}

\DeclareMathOperator{\im}{Im}
\DeclareMathOperator{\sgn}{sgn}
\DeclareMathOperator{\Ai}{Ai}

\DeclareMathOperator{\erfc}{erfc}

\newcommand{\vf}{\varphi}

\newcommand{\s}{\sigma}

\newcommand{\al}{\alpha}
\newcommand{\be}{\beta}
\newcommand{\ga}{\gamma}

\newcommand{\de}{\delta}

\newcommand{\la}{\lambda}

\newcommand{\spx}{\mathbf{x}}
\newcommand{\spy}{\mathbf{y}}

\newcommand{\spe}{\mathbf{e}}

\newcommand{\N}{\mathbb{N}}

\begin{document}
\allowdisplaybreaks[4]
\frenchspacing
\setlength{\unitlength}{1pt}

\title{{\Large\textbf{Probability of radiation of twisted photons in the infrared domain}}}

\date{}

\author{O.V. Bogdanov${}^{1),2)}$\thanks{E-mail: \texttt{bov@tpu.ru}},\; P.O. Kazinski${}^{1)}$\thanks{E-mail: \texttt{kpo@phys.tsu.ru}},\; and G.Yu. Lazarenko${}^{1)}$\thanks{E-mail: \texttt{lazarenko.georgijj@icloud.com}}\\[0.5em]
{\normalsize ${}^{1)}$ Physics Faculty, Tomsk State University, Tomsk 634050, Russia}\\
{\normalsize ${}^{2)}$ Tomsk Polytechnic University, Tomsk 634050, Russia}}

\maketitle

\begin{abstract}

The infrared asymptotics of probability of radiation of twisted photons in an arbitrary scattering process of quantum electrodynamics (QED) in a vacuum is investigated. This asymptotics is universal and corresponds to the radiation produced by a classical current. Such a radiation is known as the edge radiation. We represent it in terms of the twisted photons: the exact analytical formulas for the average number of radiated twisted photons are derived. We find the average projection of the total angular momentum of the edge radiation and the angular momentum per photon. It is shown that the edge radiation can be used as a source of twisted photons with large angular momentum. Moreover, this radiation can be employed as a superradiant coherent source of twisted photons in the infrared domain, in particular, in the THz part of the electromagnetic spectrum. Several general selection rules for the radiation and absorbtion of twisted photons are proved. These selection rules allow one, in particular, to modulate the one-particle radiation probability by means of scattering of charged particles on symmetrically arranged crystals.

\end{abstract}

\section{Introduction}

The infrared asymptotics of radiation of photons in a vacuum is known to be universal for any quantum electrodynamic (QED) process (see, e.g., \cite{BlNord37,Nords37,AkhBerQED,BolDavRok,PeskSchro,WeinbergB.12,AkhShul}). It is described by the radiation produced by the classical current of free charged particles in the in- and out-states, i.e., by the current of charged particles moving uniformly along straight lines with a break. This radiation is called the edge radiation \cite{Bossart79,Coisson,NMME80,BMNF81,AlfBash,Bess81,BTFarc,Bord.1,GKSSchYu1,GKSSchYu2,SynchRad2015}. The edge radiation can be routinely generated on the acceleration facilities (see, e.g., \cite{SmHiYo,Roy2000,Carr02,ShSHM,Mueller05,AGKSMG,Wen16}). It is used now as a brilliant source of the infrared radiation \cite{AGKSMG,Wen16,Mueller05,Carr02} and applied to the direct control of numerous low-energy excitations in solids and molecules \cite{KamTanNel,Mittle13}, to the study of mechanisms of the high-temperature superconductivity \cite{DCFZhetal13}, to THz spectroscopy and microscopy, medicine, biology, and even to the preservation of cultural heritage \cite{SynchRad2015,Mittle17}.

Surprisingly, this type of radiation has not been examined as a possible source of the vortex electromagnetic radiation, which also finds a wide range of applications in fundamental science and technology \cite{PadgOAM25,AndBabAML,TorTorTw,AndrewsSLIA}. The vortex radiation, i.e., the electromagnetic waves carrying an angular momentum, can be rigorously described in terms of the so-called twisted photons \cite{TorTorTw,MolTerTorTor,Ivanov11,JenSerprl,JenSerepj,GottfYan,JaurHac,BiaBirBiaBir}. These photons are the excitations of a quantum electromagnetic field with definite the energy, the longitudinal projection of momentum, the projection of the total angular momentum, and the helicity. The angular momentum can be regarded as an additional degree of freedom of the field and can be used for the high-density information transfer or for the manipulation of the rotational degrees of freedom of an irradiated object. The appreciation of the utility of such an instrument resulted in a rapid growth in research on the generation and application of the vortex fields, not only the electromagnetic ones but with the spin one-half as well (see, for review, \cite{BliokhVErev,LBThY}). Using recently developed general formalism \cite{BKL2}, we shall give the complete description of the edge radiation in terms of the twisted photons and find the conditions when the edge radiation can be regarded as the source of the electromagnetic waves with a considerable angular momentum. Since the edge radiation is infrared, it provides an exceptional opportunity for creation of a superradiant coherent source of twisted photons. Its intensity and, consequently, the angular momentum of radiation produced by it are proportional to the square of the particle number in the bunch of charged particles. As for the plane-wave photons, their coherent edge radiation was already observed experimentally \cite{AGKSMG,Wen16,Mueller05,Carr02}.

In \cite{BKL2}, the general formula for the probability of radiation of a twisted photon by a classical current was obtained. The replacement of the quantum current operator by a classical quantity is justified only when the quantum recoil experienced by the source in the process of radiation can be neglected. The infrared limit described by the edge radiation provides an optimal situation for such an approximation works. The definition of the infrared domain where the edge radiation dominates depends on the process at hand. Loosely speaking, the so-defined infrared domain corresponds to the photon energies much less than the minimal characteristic energy scale of the process. As we shall see, the edge radiation may extend to the X-ray range for certain processes. Of course, the coherence properties of radiation become worse in increasing the photon energy. As the particular case for application of general formulas, we shall consider the edge radiation created in the process of scattering of electrons (positrons) by crystals: the so-called volume reflection and the volume capture \cite{Andreev82,TarVor87,Scand09pi,Scand09pr,GuiBandTikh,Mazzolari14,Wistis16,Wistis17,Sytov17}. In this case, the electrons with the energies of the order $5$ GeV produce the edge radiation with the photon energies up to $850$ eV with large angular momentum.

The list of the main results obtained in this paper is presented in Conclusion. So we do not dwell on it here. In Sec. \ref{Gener_Form}, we recapitulate the main formulas of \cite{BKL2} regarding the probability of radiation of twisted photons by classical currents and prove the general selection rules for symmetric sources. Notice that the probability of radiation of twisted photons not only reveals the content of radiation in terms of the twisted photons but it can be directly observed \cite{LPBFAC,BLCBP,SSDFGCY,LavCourPad}. In Sec. \ref{IfraRed}, we find the spectrum domain where the edge radiation dominates and show that the edge radiation is a universal infrared asymptotics of any radiation of charged particles in a vacuum, the effects of a chamber \cite{Bosch02,GKSSchYu1,GKSSchYu2} where the process is evolving being supposed to be negligible. Section \ref{Processes} is devoted to the edge radiation of twisted photons. In Sec. \ref{Process_OnAx}, we consider the edge radiation produced by the charged particles with the trajectories possessing the break on the detector axis. In this case, we derive the exact formula for the probability of radiation of twisted photons and find the main characteristics of the twist of radiation: the differential asymmetry, the average projection of the total angular momentum of radiation, and the angular momentum per photon. In particular, we show that the photons with large angular momentum can be generated only by the scattering of ultrarelativistic charged particles with the sufficiently large velocity component transversal to the detector axis. We also establish there the symmetry property of the probability of radiation of twisted photons. This symmetry property holds for the radiation of twisted photons in any QED process in the far infrared. In particular, we obtain the infrared asymptotics of radiation of twisted photons produced in scattering of ultrarelativistic electrons by crystals positioned exactly on the detector axis. In Sec. \ref{Process_OffAx}, we investigate the edge radiation of twisted photons by the charged particles moving along the trajectories with the break located out of the detector axis. We find the probability of radiation of twisted photons and the angular momentum per photon of the edge radiation. The edge radiation of twisted photons for several processes of scattering of electrons by the crystals positioned out of the detector axis is studied numerically. We show that the optimal situation for the  production of twisted photons with large angular momentum in this configuration occurs when the scattered electrons possess a large angular momentum with respect to the detector axis. As another example, we consider in this section the scattering of electrons by the crystal located on the detector axis in the magnetic field of a solenoid, the axis of the solenoid coinciding with the detector axis. For such a configuration, the electrons initially moving along the detector axis escape the magnetic field of the solenoid with a large angular momentum with respect to this axis. Therefore, one should expect that the radiation produced possesses a large angular momentum too. We show that this is indeed the case and find the main characteristics of the twist of the radiation created. We also find the general bound on the maximal magnitude of the projection of the angular momentum of a twisted photon radiated by a classical current and verify this bound on several examples.

We use the system of units such that $\hbar=c=1$ and $e^2=4\pi\al$, where $\al$ is the fine structure constant.

\section{General formulas}\label{Gener_Form}

Let us briefly recall in this section the general formulas describing the radiation of twisted photons by classical currents \cite{BKL2}. Consider the theory of a quantum electromagnetic field interacting with a classical current $j_\mu(x)$. We suppose that $j_\mu(x)$ is the current density of a point charge
\begin{equation}\label{current_class}
\begin{split}
    j^\mu(x)=\,&e\Big\{\int_{\tau_1}^{\tau_2}d\tau\dot{x}^\mu(\tau)\de^4(x-x(\tau)) +\frac{\dot{x}^\mu(\tau_2)}{\dot{x}^0(\tau_2)}\theta(x^0-x^0(\tau_2))\de\big[\spx-\spx(\tau_2)-(x^0-x^0(\tau_2))\dot{\spx}(\tau_2)/\dot{x}^0(\tau_2)\big]+\\
    &+\frac{\dot{x}^\mu(\tau_1)}{\dot{x}^0(\tau_1)}\theta(x^0(\tau_1)-x^0)\de\big[\spx-\spx(\tau_1)-(x^0-x^0(\tau_1))\dot{\spx}(\tau_1)/\dot{x}^0(\tau_1)\big] \Big\},
\end{split}
\end{equation}
where $x^0(\tau_1)=-\tau_0/2$ and $x^0(\tau_2)=\tau_0/2$, and $\tau_0$ is the time period when the particle moves with acceleration. The expression \eqref{current_class} can be cast into the standard form
\begin{equation}
    j^\mu(x)=e\int_{-\infty}^{\infty}d\tau\dot{x}^\mu(\tau)\de^4(x-x(\tau)),
\end{equation}
where it is assumed that, for $\tau<\tau_1$, the particle moves with the constant velocity $\dot{x}^\mu(\tau_1)$, while, for $\tau>\tau_2$, it moves with the constant velocity $\dot{x}^\mu(\tau_2)$. On performing the Fourier transform,
\begin{equation}\label{current_fourier_gen}
    j^\mu(x)=:\int\frac{d^4k}{(2\pi)^4}e^{ik_\nu x^\nu}j^\mu(k),
\end{equation}
the last two terms in \eqref{current_class} correspond to the boundary terms in
\begin{equation}\label{current_fourier}
    j_\mu(k)=e\Big(\int_{\tau_1}^{\tau_2}d\tau \dot{x}_\mu e^{-ik_\nu x^\nu(\tau)}-\frac{i\dot{x}_\mu}{k_\la\dot{x}^\la}e^{-ik_\nu x^\nu}\Big|_{\tau_1}^{\tau_2}\Big).
\end{equation}
These boundary contributions are responsible for the radiation created by a particle when it enters to and exits from the external field (see, e.g., \cite{Bashm92,Kim96,Bosch98,Bosch02,SmHiYo,Roy2000,Carr02,ShSHM,Mueller05,AGKSMG,Wen16,Bossart79,Coisson,NMME80,BMNF81,AlfBash,Bess81,BTFarc,Bord.1,GKSSchYu1,GKSSchYu2,SynchRad2015,BolDavRok,AkhShul}). It is the so-called edge radiation. These terms gives the leading contribution to the probability of radiation of photons with small energies (see, e.g., \cite{BlNord37,Nords37,AkhBerQED,BolDavRok,PeskSchro,WeinbergB.12,AkhShul,Bord.1}). In fact, we shall study the contribution of these terms to the radiation of twisted photons. The model of a classical source producing the twisted photons is a good one when the quantum recoil experienced by the source can be neglected. The infrared regime of radiation that we are going to investigate represents an ideal situation where such an approximation works.

Let us introduce the right-handed orthonormal triple $\{\spe_1,\spe_2,\spe_3\}$ and
\begin{equation}\label{spin_eigv}
    \mathbf{e}_\pm:=\mathbf{e}_1\pm i\mathbf{e}_2.
\end{equation}
The basis vector $\spe_3$ is directed along the axis of the detector that records the twisted photons. It is clear that
\begin{equation}
    (\spe_{\pm},\spe_{\pm})=0,\qquad (\spe_{\pm},\spe_{\mp})=2,\qquad\spe^*_\pm=\spe_\mp,
\end{equation}
and arbitrary vector can be decomposed in this basis as
\begin{equation}\label{psi_decomp}
    \spx=\frac12(x_-\spe_++x_+\spe_-)+x_3\spe_3,
\end{equation}
where $x_\pm=(\spx,\spe_\pm)$ and $x_3=(\spx,\spe_3)$.

In the presence of a classical current, the process of radiation of photons is possible
\begin{equation}\label{0gX}
    0\rightarrow \ga_\al+X,
\end{equation}
where $0$ is the vacuum state, $\ga_\al$ denotes the photon in the state $\al$ recorded by the detector, and $X$ is for the other created photons. The probability of the inclusive process \eqref{0gX} is given by
\begin{equation}
    w_{incl}(\al;0)=1-e^{-n(\al;0)}\approx n(\al;0),
\end{equation}
where $n(\al;0)$ is the average number of photons in the state $\al$ created by the current $j_\mu(x)$ during the whole observation period, and it is supposed in the approximate equality that the population of the state $\al$ is small. It is always small provided the volume of the chamber where the photons are created is sufficiently large.

The density of the average number of twisted photons created by the current \eqref{current_class} reads \cite{BKL2}
\begin{multline}\label{probabil}
    dP(s,m,k_3,k_\perp)=e^2\bigg|\int d\tau e^{-i[k_0x^0(\tau)-k_3x_3(\tau)]}\Big\{\frac12\big[\dot{x}_+(\tau)a_-(s,m,k_3,k_\perp;\spx(\tau))+\\
    +\dot{x}_-(\tau)a_+(s,m,k_3,k_\perp;\spx(\tau)) \big]
    +\dot{x}_3(\tau)a_3(m,k_\perp;\spx(\tau))\Big\} \bigg|^2 \Big(\frac{k_\perp}{2k_{0}}\Big)^{3}\frac{dk_3dk_\perp}{2\pi^2},\quad k_\perp:=\sqrt{k_0^2-k_3^2},
\end{multline}
where $k_0\geq|k_3|$ is the photon energy, all the vectors components are defined as in \eqref{psi_decomp}, and the notation has been introduced \cite{BiaBirBiaBir,JaurHac,GottfYan,Ivanov11,JenSerepj,JenSerprl,BKL2}
\begin{equation}\label{mode_func_an}
\begin{split}
    a_3(m,k_\perp;\spx)&=\frac{x_+^{m/2}}{x_-^{m/2}}J_m(k_\perp x_+^{1/2}x_-^{1/2})=:j_m(k_\perp x_+, k_\perp x_-),\\
    a_\pm(s,m,k_3,k_\perp;\spx)&=\frac{ik_\perp}{sk_0\pm k_3}j_{m\pm1}(k_\perp x_+, k_\perp x_-),
\end{split}
\end{equation}
where $s=\pm1$ is the photon helicity and $m\in \mathbb{Z}$ is the projection of the total angular momentum of a photon onto the detector axis. The functions $j_m(p,q)$ are entire functions of the complex variables $p$ and $q$. It is also assumed in \eqref{probabil} that the origin of the reference frame is taken on the line passing through the detector axis. The expression \eqref{probabil} is invariant under the translations of the origin of the system of coordinates along the detector axis.

Formula \eqref{probabil} is obviously written for the current density of $N$ charged particles and for distributed currents. We shall also need the following general statement. Let the distributed current density $j^i(t,\spx)$ be invariant under the rotation by an angle of $2\pi/r$, $r\in\N$, around the detector axis for all $t$. Then
\begin{equation}\label{selection_rule}
    dP(s,m,k_3,k_\perp)=0,\qquad m\neq lr,\quad l\in \mathbb{Z}.
\end{equation}
A similar property is known to be fulfilled for the scattering of twisted photons by microstructures \cite{RubDun08,RubDun11}. In order to prove this statement, we employ the pictorial representation of the radiation amplitude of twisted photons (see, for details, \cite{BKL2}) in terms of the plane-wave ones. Let us single out the circle that is obtained by the rotation of some point $\spx$ around the detector axis. In accordance with the prescription described in \cite{BKL2}, the contribution to the radiation amplitude of every point on this circle comes with the common phase factor
\begin{equation}
    e^{im\vf}+e^{im(\vf+2\pi/r)}+e^{im(\vf+4\pi/r)}+\cdots+e^{im(\vf+2\pi(r-1)/r)}=re^{im\vf}\de_{m,lr},
\end{equation}
which was to be proved. Another way to prove this statement is to see that the radiation amplitude entering into \eqref{probabil} acquires the phase factor $e^{im\vf}$ when the rotation by an angle of $\vf$ is performed around the detector axis (see \cite{BKL2}). This immediately implies that either $2\pi m/r=2\pi l$ or the amplitude vanishes. In particular, if the current density $j^i(t,\spx)$ is invariant with respect to the rotations by an arbitrary angle, which formally corresponds to $r\rightarrow\infty$, then the average number of radiated twisted photons is nonzero for $m=0$ only. Here, of course, we assume that the radiation fields created by $j^i(t,\spx)$ add up coherently.

The transformation property of the amplitude can be used to show another one statement. Let the current density $j^i(t,\spx)$ of $r$ identical charged particles be obtained from the current density of one charged particle by rotating its trajectory by an angle of $2\pi k/r$, $k=\overline{0,r-1}$, around the detector axis. Then
\begin{equation}\label{probabil_sym}
    dP(s,m,k_3,k_\perp)=r^2dP_1(s,m,k_3,k_\perp)\de_{m,lr},\qquad l\in \mathbb{Z},
\end{equation}
where $dP_1(s,m,k_3,k_\perp)$ is the average number of twisted photons created by the current density of one charged particle. For example, if one considers the ideal situation when $r$ electrons move in the helical undulator along the ideal circular helix such that their trajectories pass to each other under the rotation around the detector axis by an angle of $2\pi/r$, then the average number of photons is proportional to $\de_{m,lr}$, $l\in \mathbb{Z}$. On the other hand, the forward radiation of twisted photons obeys in this case the selection rule $m=\pm n$, where $n$ is the harmonic number and the sign is determined by the handedness of the helix \cite{SasMcNu,BHKMSS,TaHaKa,Rubic17,BKL2}. Consequently, the average number of twisted photons is nonzero only when $n=lr$ (see \cite{Bord.1}).

Formula \eqref{probabil_sym} is the particular case of a more general statement on the form of the average number of twisted photons produced by the system of identical charged particles moving along the trajectories that are obtained from each other by the rotation around the detector axis, the translation along it, and the translation in time. Namely, consider a set of trajectories of the identical charged particles that are obtained from one trajectory by the rotation by an angle of $\vf_k$, the translation along $\spe_3$ by $x_3^k$, and the translation in time $x_0^k$. Then, as follows from the transformation properties of the integrand of \eqref{probabil}, the average number of twisted photons radiated by such a system of particles can be cast into the form
\begin{equation}\label{probabil_sym1}
    dP(s,m,k_3,k_0)=\Big|\sum_{k=1}^r e^{im\vf_k+ik_3x_3^k-ik_0x_0^k}\Big|^2dP_1(s,m,k_3,k_0)=:I(m,k_3,k_0)dP_1(s,m,k_3,k_0),
\end{equation}
where $r$ is the number of particles. In particular, if
\begin{equation}\label{symm_transf}
    \vf_k=\frac{2\pi k}{r},\qquad x_3^k=\frac{\la_0}{2\pi} \vf_k,\qquad x_0^k=\frac{\la_0}{2\pi\be_\parallel} \vf_k,
\end{equation}
where $\la_0$ and $\be_\parallel$ are some fixed parameters, then
\begin{equation}\label{interfer_factor}
    I(m,k_3,k_0)=\frac{\sin^2(\pi\de)}{\sin^2(\pi\de/r)},\qquad\de:=m+\frac{k_0\la_0}{2\pi}(\be_\parallel^{-1}-n_3),
\end{equation}
where $n_3:=k_3/k_0$. This interference factor modulates the one-particle radiation probability. It possesses the sharp global maxima at
\begin{equation}
    \de=lr,\qquad l\in \mathbb{Z},
\end{equation}
where $I(m,k_3,k_0)=r^2$, and the lateral local maxima, where $I(m,k_3,k_0)\sim1$. The function $I(m,k_3,k_0)$ vanishes at $\de\in \mathbb{Z}$ when $\de\neq lr$. Therefore, for $\la_0=0$, we reproduce \eqref{probabil_sym}. In the general case, we have the selection rule
\begin{equation}\label{supersel_rule}
    m=\sgn(\la_0)n+lr,\qquad k_0=k_0^n:=\frac{2\pi n}{|\la_0|(\be_\parallel^{-1}-n_3)},\qquad n\in \mathbb{Z}\setminus\{0\},
\end{equation}
at the maxima. For the photon energy $k_0=k_0^n$, the average number of radiated twisted photons vanishes for those $m$ that do not satisfy \eqref{supersel_rule}. If $\be_\parallel\in[0,1)$ then the harmonic number $n=\overline{1,\infty}$. Introducing the notation
\begin{equation}
    \omega_0:=2\pi\be_\parallel/|\la_0|,
\end{equation}
we see that
\begin{equation}
    k_0^n=\frac{\omega_0n}{1-\be_\parallel n_3},
\end{equation}
i.e., we have exactly the spectrum of the forward radiation of twisted photons by the helical undulator \cite{BKL2}. The case $r\rightarrow\infty$ formally corresponds to the scattering of particles on the spiral phase plate commonly used to produce the twisted electrons \cite{BliokhVErev,LBThY} and photons \cite{twpxray} or to the helical trajectory of a charged particle in the undulator, for example, (see [Fig. 5, \cite{BKL2}]). In this case, we obtain the selection rule
\begin{equation}\label{supersel_rule1}
    m=\sgn(\la_0)n,
\end{equation}
i.e., all the twisted photons radiated at the $n$th harmonic possess the total angular momentum \eqref{supersel_rule1}. We should mention once again that all these selection rules imply the coherent addition of radiation fields of charged particles.

The above general statements are valid for the inverse process too, i.e., the system possessing such a current density does not absorb the twisted photons that do not obey these selection rules within the bounds of the approximations made in replacing the current operator by the classical quantity. This property follows from unitarity of a quantum evolution.

\section{Infrared asymptotics}\label{IfraRed}

In order to find the infrared asymptotics of \eqref{probabil}, we parameterize the worldline by the laboratory time, $\tau=t=x^0$, and assume that the trajectory of a charged particle has the form
\begin{equation}\label{asymptotes}
    \spx(t)=\spx_0+\mathbf{v}t+\de\spx(\omega t)=\spy_0+\mathbf{u}t+\de\spy(\omega t),
\end{equation}
where $\spx_0$, $\spy_0$, $\mathbf{v}$, and $\mathbf{u}$ are constant vectors, $\omega$ characterizes the time scale of variations of the trajectory, and
\begin{equation}
\begin{gathered}
    \de\spx(\tau)\underset{\tau\rightarrow+\infty}{\rightarrow}0,\qquad \de\spy(\tau)\underset{\tau\rightarrow-\infty}{\rightarrow}0,\\
    \de\dot{\spx}(\tau)\underset{\tau\rightarrow+\infty}{\rightarrow}0,\qquad \de\dot{\spy}(\tau)\underset{\tau\rightarrow-\infty}{\rightarrow}0.
\end{gathered}
\end{equation}
The parameters $\spx_0$, $\spy_0$, $\mathbf{v}$, and $\mathbf{u}$ specify the asymptotes of the trajectory in the future and in the past. More precisely, we suppose that
\begin{equation}\label{IR_cond1}
    k_3|\de x_3(\tau)|\ll1,\qquad k_\perp |\de x_+(\tau)|\ll1,
\end{equation}
when $\tau\gtrsim 2\pi$, and the same estimates hold for $\de\spy(\tau)$. Then, we partition the integral over $t$ in \eqref{probabil} into two,
\begin{equation}\label{int_part}
    \int_{-\infty}^\infty d t\cdots=\int_{-\infty}^0 d t\cdots+\int_{0}^\infty d t\cdots,
\end{equation}
and stretch the integration variable
\begin{equation}
    t\rightarrow\frac{t}{k_0(1-n_3 v_3)},\qquad t\rightarrow\frac{t}{k_0(1-n_3 u_3)},
\end{equation}
in the first and the second integrals, respectively. Hereinafter, $n_3:=k_3/k_0$, $n_\perp:=k_\perp/k_0$, and so
\begin{equation}
    n_3^2+n_\perp^2=1.
\end{equation}
Having performed such a transform, we see from \eqref{IR_cond1} that, when \cite{BolDavRok}
\begin{equation}\label{IR_cond2}
    \frac{\omega}{k_0(1-n_3 v_3)}\gtrsim 2\pi,\qquad \frac{\omega}{k_0(1-n_3 u_3)}\gtrsim 2\pi,
\end{equation}
and the estimates \eqref{IR_cond1} are satisfied, the particle trajectory in the partitioned integral \eqref{int_part} can be replaced, with good accuracy, by the corresponding asymptotes in the future and in the past. The whole trajectory in this approximation is discontinuous both in the velocity and the position of a charged particle with the discontinuity point $t=0$:
\begin{equation}\label{IR_traj}
    \spx(t)=\left\{
              \begin{array}{ll}
                \spx_0+\mathbf{v}t, & \hbox{$t>0$;} \\
                \spy_0+\mathbf{u}t, & \hbox{$t<0$.}
              \end{array}
            \right.
\end{equation}
For such a trajectory, the second integral in \eqref{int_part} is obtained from the first one by a change of sign and the replacement $\spx_0\rightarrow\spy_0$, $\mathbf{v}\rightarrow\mathbf{u}$. Hence, in evaluating the integrals it is sufficient to consider the first integral.

This integral possesses the physical meaning in itself (see, e.g., \cite{Bess81}). It describes the amplitude of the twisted photon production in the processes of the ``instantaneous'' stopping of a charged particle in a target and the ``instantaneous'' acceleration of a charged particle from a state of rest \cite{BolDavRok}. The latter process is realized, for example, in the production of charged particles in nuclear reactions. In these cases, the contribution of one of the integrals in \eqref{int_part} is zero.

In the ultrarelativistic regime, the main part of radiation is concentrated in the cone with the opening of the order $\ga^{-1}$. Therefore,
\begin{equation}
    n_3\approx1-\frac{n_\perp^2}{2},\qquad v_3\approx1-\frac{1+\be_\perp^2\ga^2}{2\ga^2},
\end{equation}
where $\be_\perp=|v_+|$. In this case, the conditions \eqref{IR_cond1}, \eqref{IR_cond2} look as
\begin{equation}\label{IR_cond_rel}
    k_0\de x_3(2\pi)\ll1,\qquad n_kk_0\frac{K}{\ga}|\de x_+(2\pi)|\ll1,\qquad 2\pi k_0\lesssim\frac{2\omega\ga^2}{1+K^2(1+n_k^2)},
\end{equation}
where $K:=\be_\perp\ga$ and $n_k:=n_\perp/\be_\perp$. Notice that the right-hand side of the last inequality is the energy of twisted photons at the first harmonic of the forward undulator radiation \cite{BKL2} provided that $\omega$ is the circular oscillation frequency of the electron in the undulator. We see that the domain of the infrared radiation we are investigating enlarges with increasing $\omega$ and $\ga$.

Mention should be made that if the conditions \eqref{IR_cond1}, \eqref{IR_cond2}, or \eqref{IR_cond_rel}, hold simultaneously for the trajectories of several particles then the trajectories of these particles can be replaced by the straight lines with break of the form \eqref{IR_traj}, and the radiation amplitudes corresponding to these particles ought to be added up. Since the asymptotes \eqref{IR_traj} are much easier to control than the complicated dynamics in between them, it is much easier to create a coherent radiation of twisted photons in the infrared regime. The probability of such a radiation is proportional to $N^2$, where $N$ is the number of identical charged particles (see, e.g., \eqref{probabil_sym}). The condition when the radiation of different charged particles add up coherently is the same as for the radiation of plane-wave photons. In particular, the bunch of charged particles radiates coherently at a given wavelength if the bunch size is less than that wavelength.


\section{Processes}\label{Processes}
\subsection{Trajectories with a break on the detector axis}\label{Process_OnAx}

We begin with the integrals determining the contribution to the radiation amplitude of the part of the trajectory \eqref{IR_traj} for $t>0$ with $\spx_0=0$. This class of trajectories describes, in particular, the far infrared asymptotics of the average number of radiated twisted photons \eqref{probabil}. In this limit, not only are the conditions \eqref{IR_cond1}, \eqref{IR_cond2} met, but also
\begin{equation}\label{IR_cond3}
    k_3 |x_3(0)|\ll1,\qquad k_\perp |x_+(0)|\ll1,
\end{equation}
and the same estimates should be valid for $\de\spy$. Notice that, in contrast to the plane-wave photon radiation probability, the expression \eqref{probabil} is not invariant under the translations of the origin of the reference frame that are perpendicular to the detector axis. Therefore, one cannot vanish arbitrary $\spx_0$ by a proper choice of the system of coordinates.

Let us introduce the notation
\begin{equation}\label{I_int1}
\begin{split}
    I_3&:=\int_0^\infty dt v_3e^{-ik_0t(1-n_3v_3)}j_m(k_\perp v_+t,k_\perp v_-t),\\
    I_\pm&:=\frac{in_\perp}{s\mp n_3}\int_0^\infty dt v_\pm e^{-ik_0t(1-n_3v_3)}j_{m\mp1}(k_\perp v_+t,k_\perp v_-t).
\end{split}
\end{equation}
Taking into account that
\begin{equation}
    j_m(k_\perp v_+t,k_\perp v_-t)=e^{im\de}J_m(k_\perp|v_+|t),
\end{equation}
where $\de:=\arg v_+$, the integrals \eqref{I_int1} are reduced to the Laplace transform of the Bessel function \cite{GrRy}
\begin{equation}
    \int_0^\infty dxe^{-px}J_m(x)=p^{-1-m}\frac{\big(1+\sqrt{1+p^{-2}}\big)^{-m}}{\sqrt{1+p^{-2}}},\qquad m\geq0.
\end{equation}
Using this formula, we obtain
\begin{equation}\label{rad_ampl_onaxis}
\begin{alignedat}{2}
    I_3+\frac12 (I_++I_-)&=\frac{i^{-1-m}}{k_0n_\perp^2}e^{im\de}\Big(\frac{v_3-n_3}{\kappa(v)}-s\sgn(m)\Big)q^{|m|}(v),&\qquad&\text{for $|m|>1$}; \\
    I_3+\frac12 (I_++I_-)&=\frac{i^{-1}}{k_0n_\perp^2}\Big(\frac{v_3-n_3}{\kappa(v)}+n_3\Big),&\qquad&\text{for $m=0$};
\end{alignedat}
\end{equation}
where
\begin{equation}
\begin{split}
    \kappa(v)&:=\big[(1-n_3v_3)^2-n_\perp^2\be_\perp^2\big]^{1/2},\\
    q(v)&:=\frac{n_\perp\be_\perp}{1-n_3v_3+\kappa(v)}=\frac{1-n_3v_3-\kappa(v)}{n_\perp\be_\perp},\qquad 0\leq q<1.
\end{split}
\end{equation}
Notice that formula \eqref{rad_ampl_onaxis} is exact. If a charged particle moves along the detector axis, i.e., $\be_\perp=0$, then the contribution at $m=0$ only survives in \eqref{rad_ampl_onaxis}. Taking into account the contribution of the second part of the trajectory ($t<0$) with $\spy_0=0$, we have from \eqref{rad_ampl_onaxis}
\begin{equation}\label{rad_ampl_onaxis_full}
    I_3+\frac12 (I_++I_-)=\frac{i^{-1-m}}{k_\perp n_\perp}\Big[\Big(\frac{v_3-n_3}{\kappa(v)}-s\sgn(m)\Big)q^{|m|}(v)e^{im\de(v)}-(v\leftrightarrow u)\Big],
\end{equation}
for all $m$. Further we ought to square the modulus of \eqref{rad_ampl_onaxis_full} and substitute the result into \eqref{probabil}. This leads to
\begin{equation}\label{aver_num_ph}
\begin{split}
    dP(s,m,k_3,k_\perp)=&\frac{e^2}{16\pi^2} \Big\{\Big[\frac{v_3-n_3}{\kappa(v)}-s\sgn(m)\Big]^2 q^{2|m|}(v)+\\
    &+\Big[\frac{u_3-n_3}{\kappa(u)} -s\sgn(m)\Big]^2 q^{2|m|}(u) -2\cos( m\de_{12})\times\\
    &\times\Big[\frac{v_3-n_3}{\kappa(v)}-s\sgn(m)\Big] \Big[\frac{u_3-n_3}{\kappa(u)}-s\sgn(m)\Big] q^{|m|}(v)q^{|m|}(u) \Big\} \frac{dk_3dk_\perp}{k_0 k_\perp}.
\end{split}
\end{equation}
where $\de_{12}:=\de(v)-\de(u)$ is the phase difference, which is brought to the interval $(-\pi,\pi]$.

The structure of the expression \eqref{rad_ampl_onaxis} implies the symmetry property
\begin{equation}\label{symm_prop}
    dP(s,m,k_3,k_\perp)=dP(-s,-m,k_3,k_\perp)
\end{equation}
for the average number of twisted photons radiated by an \emph{arbitrary} number of charged particles with the trajectories of the form \eqref{IR_traj} with $\spx_0=\spy_0=0$ and the arbitrary velocities $\mathbf{v}$, $\mathbf{u}$. Indeed, it follows from \eqref{rad_ampl_onaxis} that the average number of twisted photons produced in this case is of the form
\begin{equation}
    dP(s,m)=\sum_{l=1}^k a_l(|m|,s\sgn(m))e^{im\de_l} \Big[\sum_{l=1}^k a_l(|m|,s\sgn(m))e^{im\de_l}\Big]^*,
\end{equation}
where $a_l((|m|,s\sgn(m)))$, $l=\overline{1,k}$, are real-valued quantities. Therefore, we obtain \eqref{symm_prop}. The symmetry property \eqref{symm_prop} also holds for the radiation produced by a charged particle moving along an arbitrary planar trajectory provided that the detector axis belongs to the orbit plane \cite{BKL2}. Contrary to that, in the case we consider here, the trajectories do not lie on one plane and the direction of the detector axis is arbitrary. The only restriction is that the breaks of the trajectories \eqref{IR_traj} are located at one point and this point lies on the detector axis. In particular, the symmetry property \eqref{symm_prop} is fulfilled in the far infrared for the radiation produced by charged particles moving along the arbitrary trajectories with asymptotes \eqref{asymptotes}.

The trajectory of the form \eqref{IR_traj} does not possess a distinguished length scale when $\spx_0=\spy_0=0$. This results in a simple dependence of \eqref{aver_num_ph} on the photon energy $k_0$. Of course, this property is a consequence of the approximations made in replacing the exact particle trajectory by the two straight lines \eqref{IR_traj}. Such a simple dependence on $k_0$ disappears when the conditions \eqref{IR_cond1}, \eqref{IR_cond2}, and \eqref{IR_cond3} become violated. In particular, when one considers the reflection of ultrarelativistic electrons (or positrons) from a crystal (see, e.g., \cite{BellMaish16,Biryuk,ShTBE,EKPT15,TarVor87,Scand09pi,Scand09pr,GuiBandTikh,Mazzolari14,Wistis16,Wistis17,Sytov17}), the trajectory of the particle can be replaced by \eqref{IR_traj} in evaluating the radiation of twisted photons provided that the photon energy satisfies the estimate
\begin{equation}\label{phot_energ_ub}
    k_0\lesssim \frac{\al^2\gamma}{\pi(1+(1+n_k^2)K^2)},
\end{equation}
where the photon energy is measured in the rest energies of the electron, $0.511$ MeV. This crude estimate follows from the last inequality in \eqref{IR_cond_rel}, where
\begin{equation}
    \omega\sim E\ga^{-1}\approx \al^2\ga^{-1}.
\end{equation}
Here the characteristic strength of the electromagnetic field is measured in the units of the critical field
\begin{equation}\label{critical_field}
    E_0=\frac{m^2}{|e|\hbar}\approx 4.41\times 10^{13}\;\text{G}=1.32\times 10^{16}\;\text{V/cm}.
\end{equation}
The typical crystalline field strengths are of the order $E\approx r_B^{-2}$, where $r_B=\al^{-1}$ is the Bohr radius in the Compton wavelengths (see, e.g., \cite{KimCue,AkhShul,BaKaStrbook}). The reflection angle in this case is of the order of the Lindhard critical angle
\begin{equation}
    \theta_c\sim (\al/\ga)^{1/2}.
\end{equation}
Therefore,
\begin{equation}
    K\lesssim(\al \ga)^{1/2}.
\end{equation}
Even larger $K$ can be achieved not for the volume reflected but for the volume captured electrons in a bent crystal (see, e.g., \cite{Sytov17,Wistis17,Wistis16,Mazzolari14,Scand09pi,Scand09pr}).

\begin{figure}[!t]
\centering
\includegraphics*[align=c,width=0.35\linewidth]{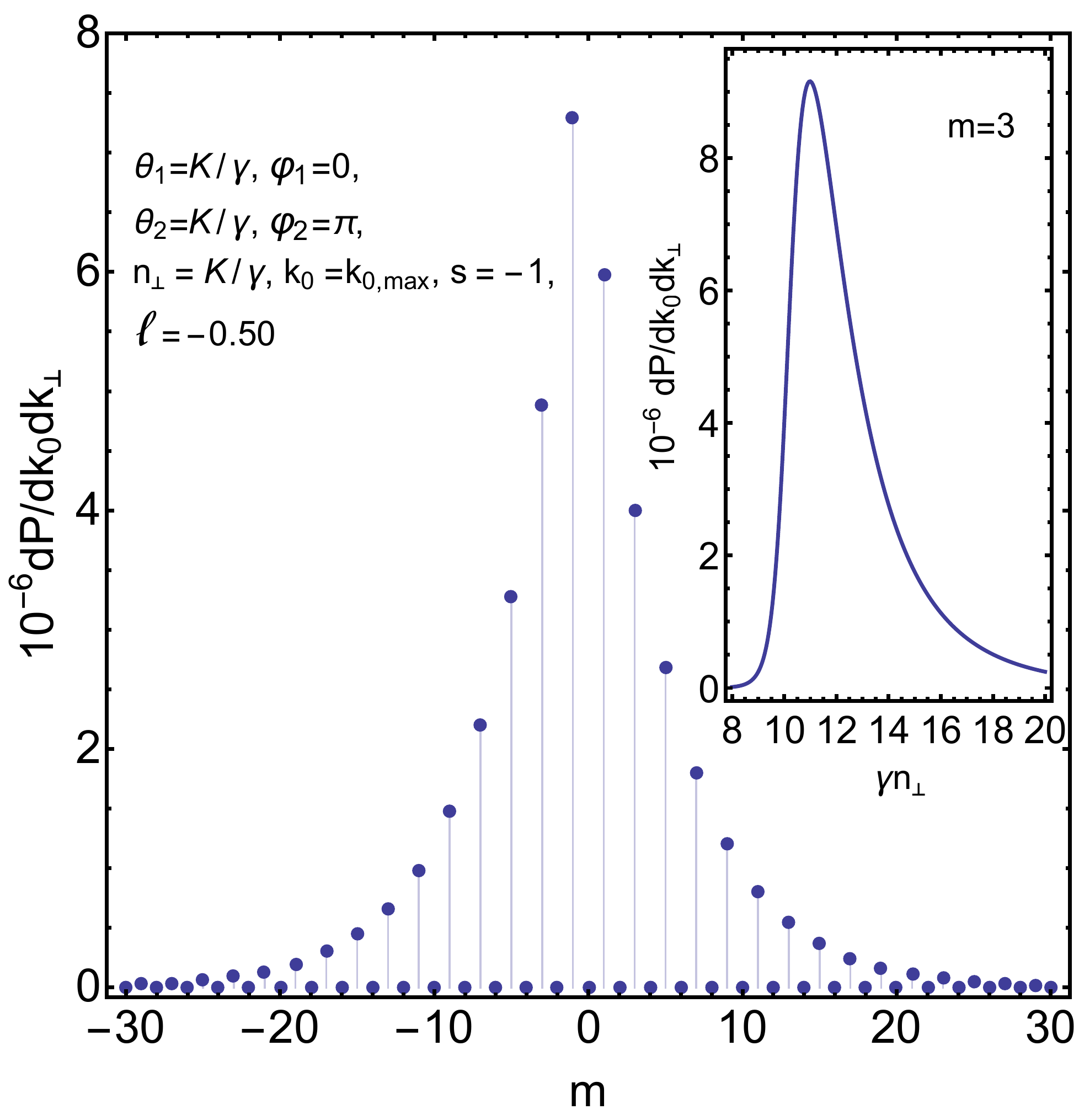}\qquad\quad
\includegraphics*[align=c,width=0.35\linewidth]{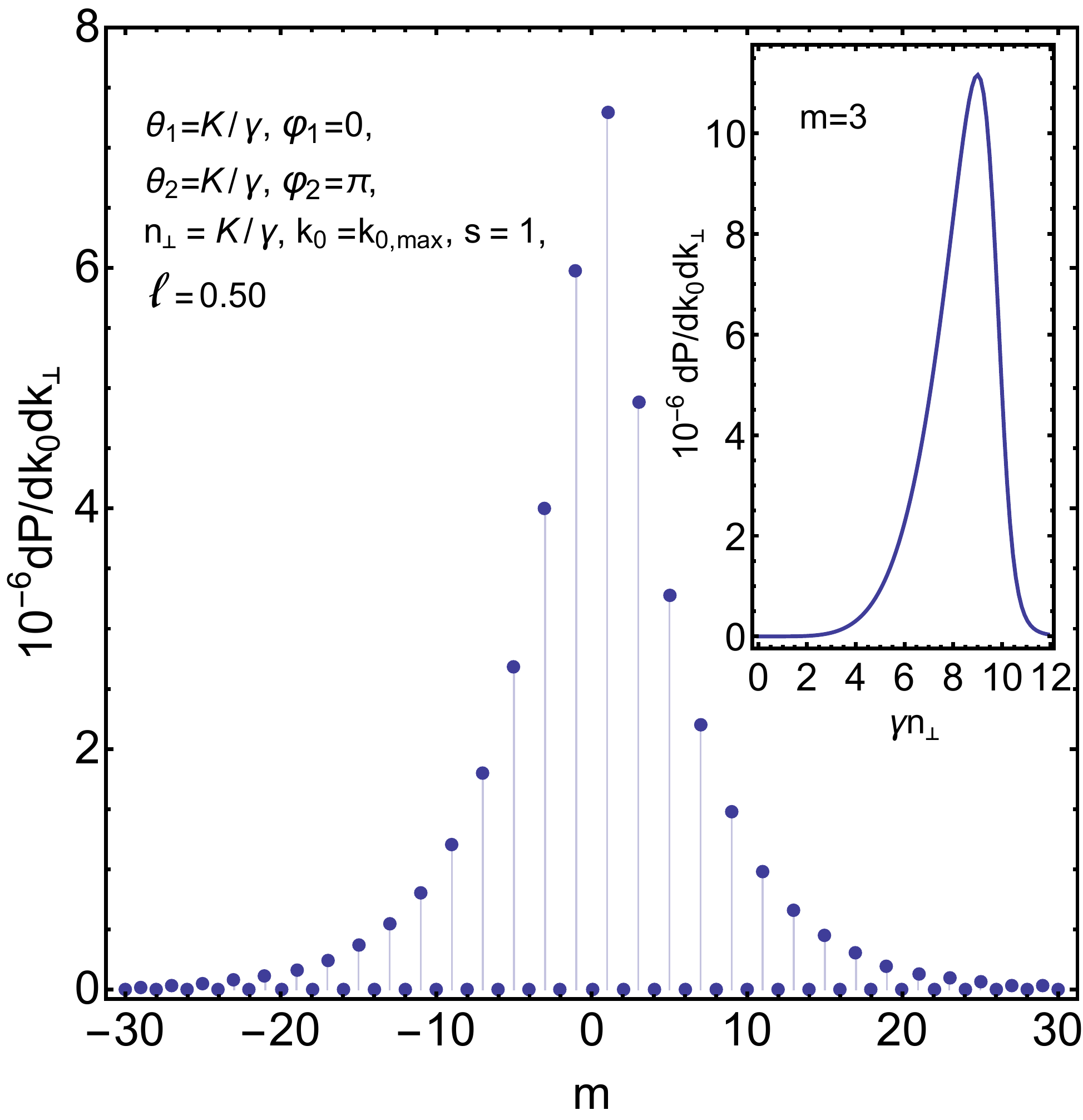}\\
\caption{{\footnotesize The average number of twisted photons against the projection of the total angular momentum produced in the elastic scattering of the electron with the Lorentz-factor $\ga=10^3$ off the target positioned on the detector axis. The photon energy $k_{0,max}:=\al^2\ga/(\pi K^2)\approx 1.7\times10^{-4}\approx 87$ eV is taken from the upper bound \eqref{phot_energ_ub}. The parameter $K:=\be_\perp\ga=10$, where $\be_\perp$ is the velocity component perpendicular to the detector axis. The angles $\theta_1$, $\vf_1$ are the polar and azimuth angles of the initial velocity of the electron with respect to the detector axis. The angles $\theta_2$, $\vf_2$ are the polar and azimuth angles of the final velocity of the electron with respect to the detector axis. The distributions over $m$ obey the symmetry property \eqref{symm_prop} and the bound \eqref{m_max}. The average number of twisted photons vanishes for even $m$. The value of the average projection of the angular momentum per photon $\ell$ is well described by \eqref{ell_onaxis}. The insets: The dependence of the average number of twisted photons on $n_\perp$ at $m=3$ and the helicities $s=\pm1$.}}
\label{on_axis2_plots}
\end{figure}

\begin{figure}[!t]
\centering
a)\;\includegraphics*[align=c,width=0.35\linewidth]{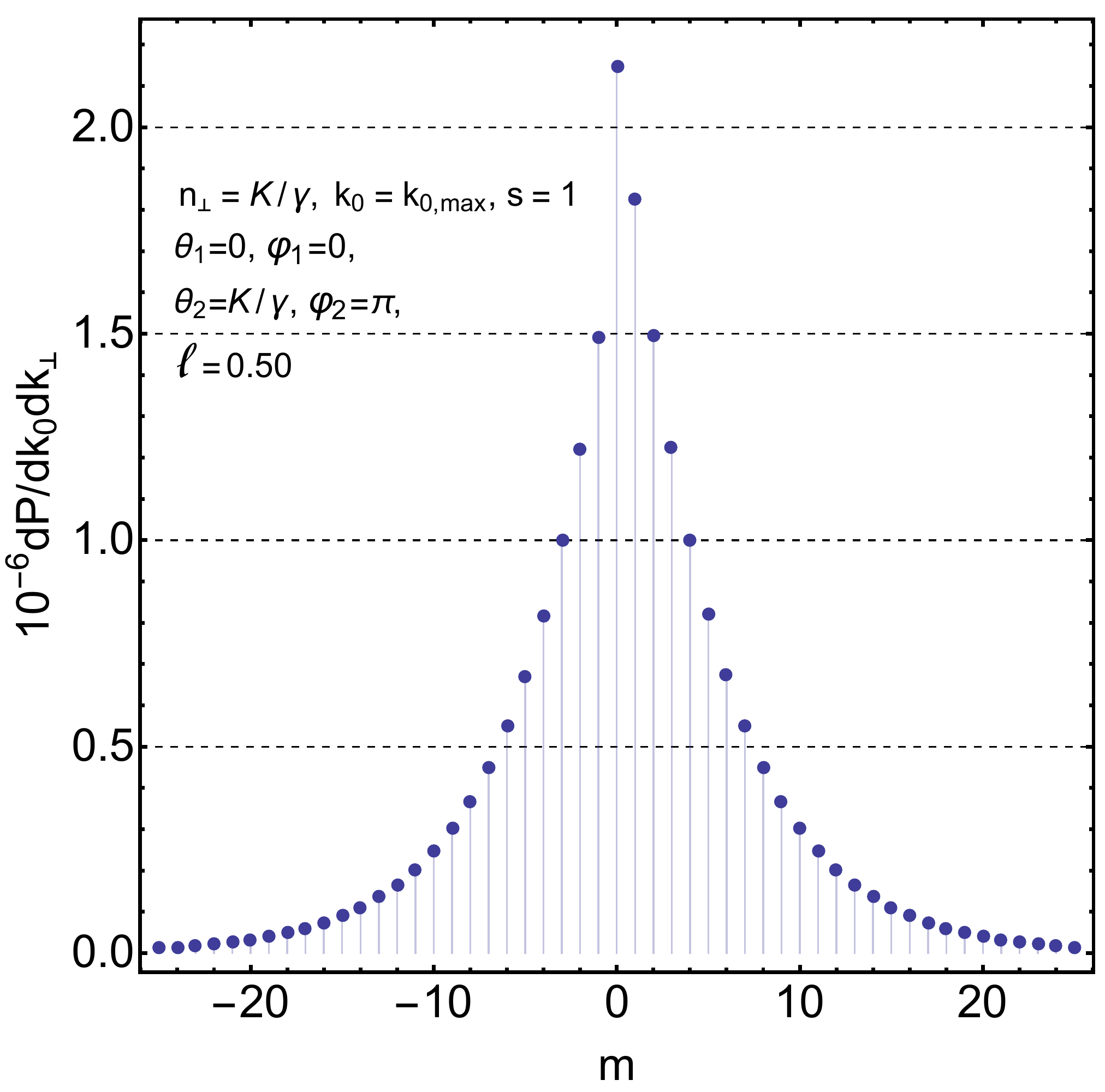}\qquad\quad
b)\;\includegraphics*[align=c,width=0.34\linewidth]{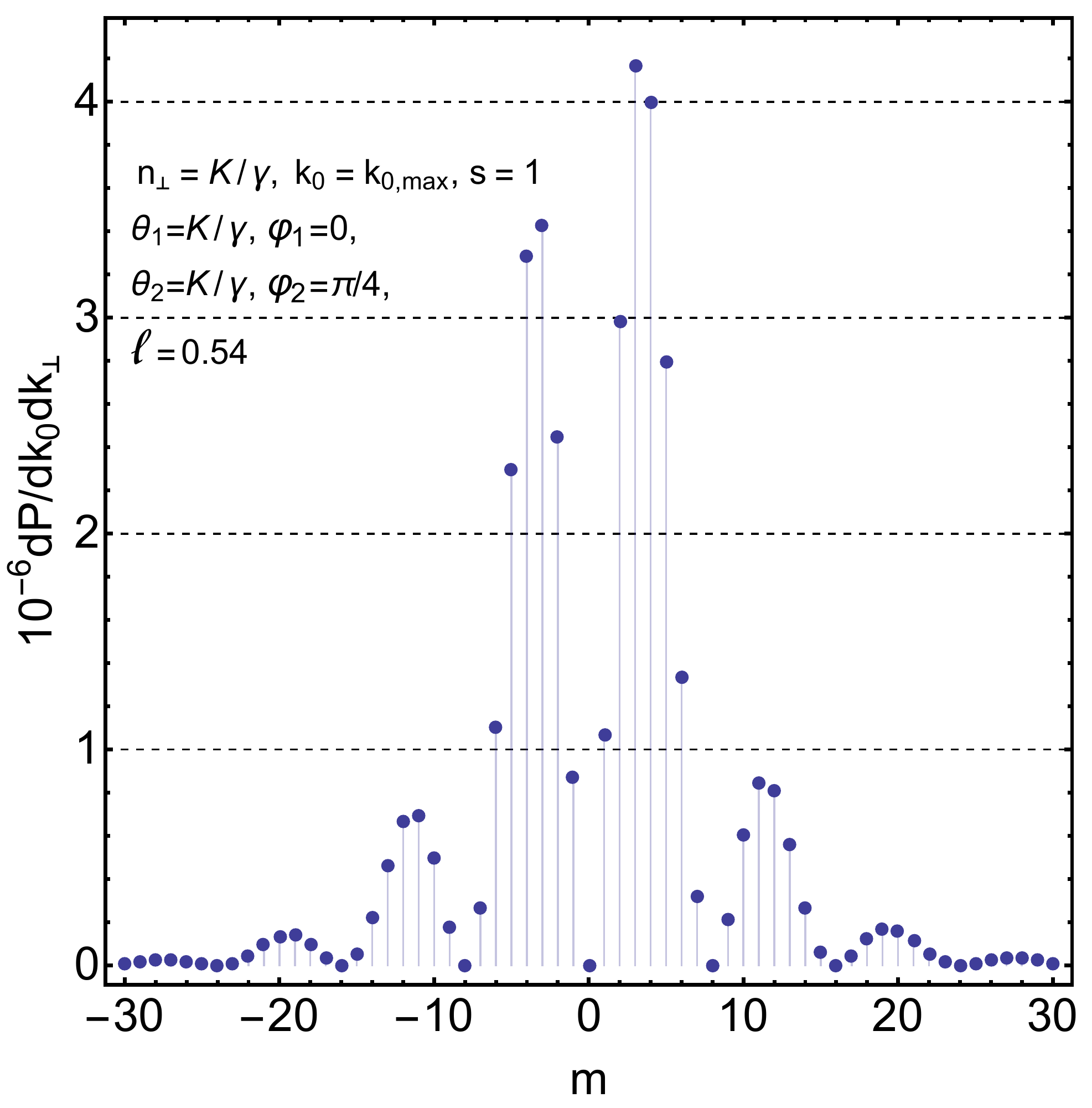}\\
c)\;\includegraphics*[align=c,width=0.35\linewidth]{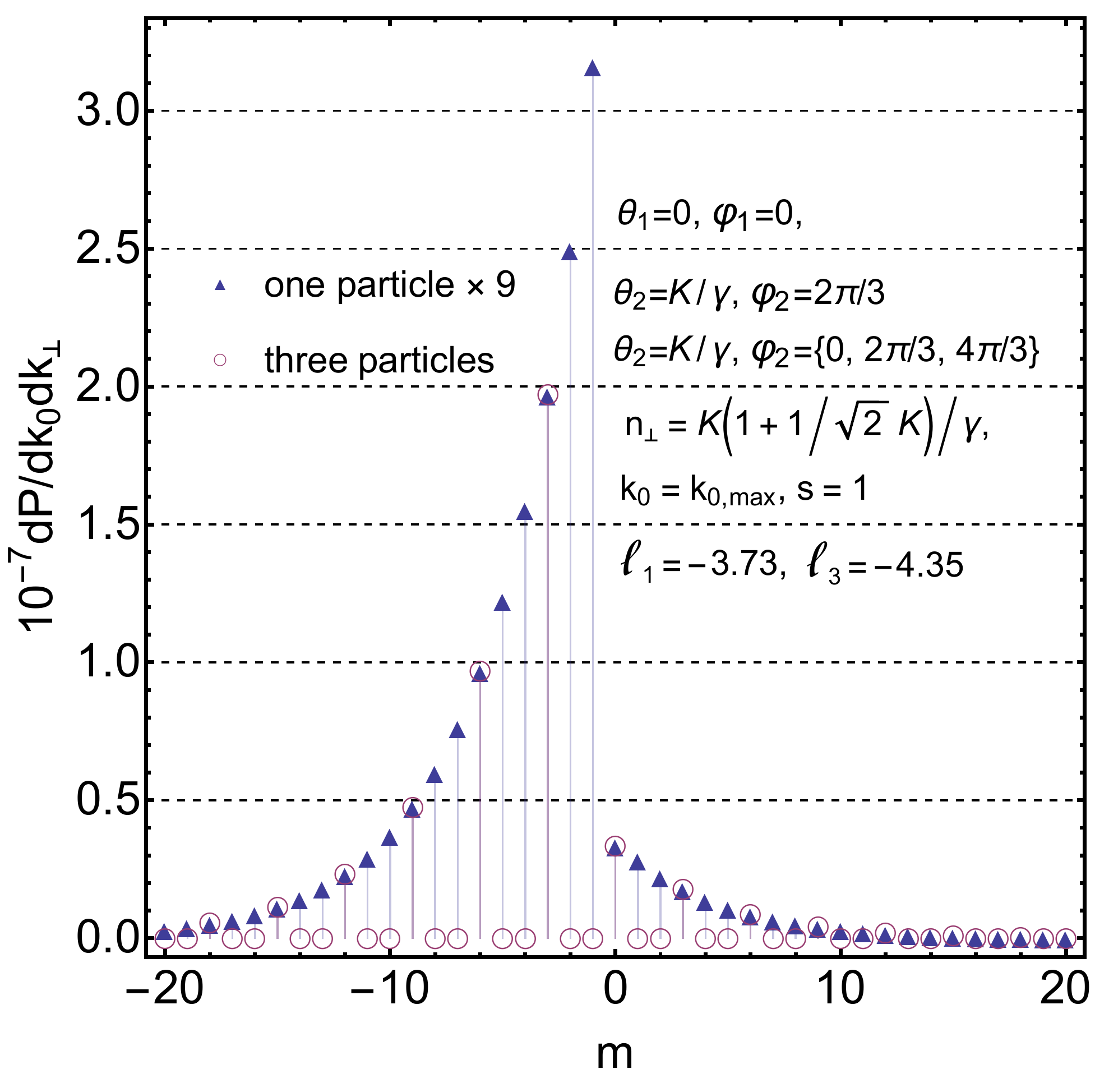}\qquad\quad
d)\;\includegraphics*[align=c,width=0.355\linewidth]{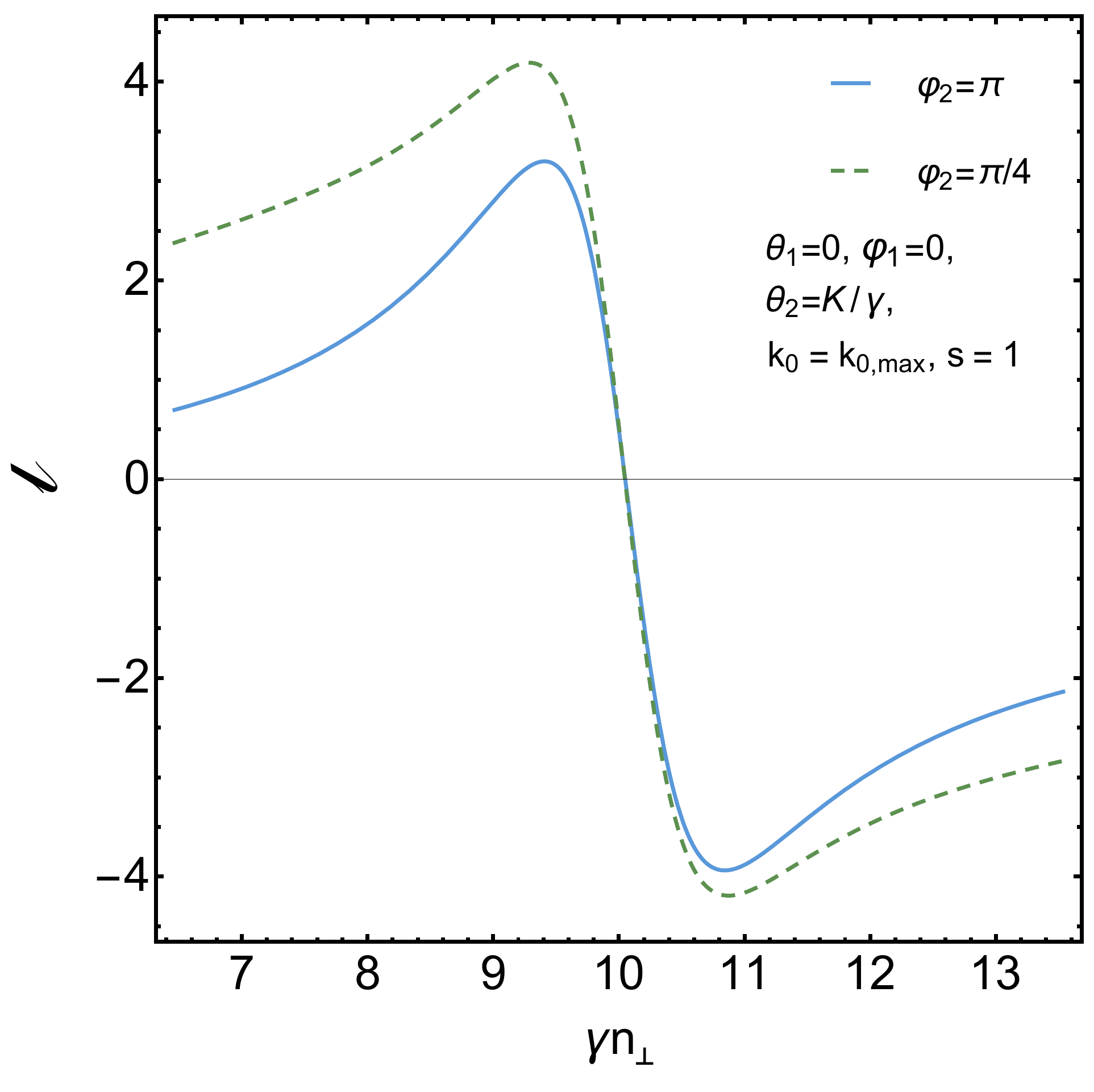}
\caption{{\footnotesize The parameters $\ga$, $K$, and $k_{0,max}$ and the definitions of the angles $\theta_{1,2}$, $\vf_{1,2}$ are the same as in Fig. \ref{on_axis2_plots}. (a-c) The average number of twisted photons against the projection of the total angular momentum produced in the elastic scattering of the electron off the target positioned on the detector axis for the different initial and final directions of the electron velocity. The distributions over $m$ obey the bound \eqref{m_max}. The values of the average projection of the angular momentum per photon $\ell$ are well described by \eqref{ell_onaxis}, \eqref{mom_per1_onax}. (b) The minima of the average number of twisted photons occur with the period \eqref{oscill_per}. (c) The production of twisted photons in scattering of one and three electrons is compared. The distributions over $m$ satisfy the property \eqref{probabil_sym}. (d) The dependence of the average projection of the angular momentum per photon on $n_\perp$. This dependence is well described by \eqref{ell_onaxis}.}}
\label{on_axis4_plots}
\end{figure}

For the processes we consider, the twisted photons with large total angular momentum can be generated only when $q\rightarrow1$. This occurs in the ultrarelativistic case, $\ga\gtrsim10$, and for $K\gtrsim3$. Then
\begin{equation}\label{q_app}
    q\approx\frac{K^2(n_k^2+1)+1-\big[K^4(n_k^2-1)^2+2K^2(n_k^2+1)+1\big]^{1/2}}{2K^2n_k} \approx1-[\de n_k^2+K^{-2}]^{1/2},
\end{equation}
where the absolute value of $\de n_k:=n_k-1$ must be much less than unity. It follows from \eqref{aver_num_ph} that the average number of twisted photons $dP(m)$ produced in this process drops by a factor of $e^2$ in comparison with $dP(0)$ for
\begin{equation}\label{m_max}
    |m|=m_{max}\approx [\de n_k^2+K^{-2}]^{-1/2}.
\end{equation}
Hereinafter we use the approximate expressions
\begin{equation}\label{approx_exprsns}
\begin{gathered}
    1-n_3v_3\approx\frac{1+K^2(1+n_k^2)}{2\ga^2},\qquad n_3-v_3\approx\frac{1+K^2(1-n_k^2)}{2\ga^2},\\
    \kappa^2(v)\approx\frac{(1+K^2(n_k-1)^2)(1+K^2(n_k+1)^2)}{4\ga^4},
\end{gathered}
\end{equation}
where only the leading terms at large $\ga$'s are kept. The typical dependence $dP(s,m)$ is given in Figs. \ref{on_axis2_plots}, \ref{on_axis4_plots}.

It is not difficult to find the main parameters characterizing the twist of the radiation (see \cite{BKL2}). For the amplitude \eqref{rad_ampl_onaxis}, i.e., for the process of the instantaneous acceleration or stopping, the average number of photons produced is given by \eqref{aver_num_ph} with $\mathbf{u}=0$. Employing the approximate expressions \eqref{approx_exprsns}, we can write
\begin{equation}\label{dP_approx}
    dP(s,m,k_3,k_\perp)\approx \frac{e^2}{16\pi^2} \Big(\frac{K^2(n_k^2-1)-1}{\big[K^4(n_k^2-1)^2+2K^2(n_k^2+1)+1\big]^{1/2}}  -s\sgn(m)\Big)^2q^{2|m|}(v) \frac{dk_3dk_\perp}{k_0 k_\perp},
\end{equation}
where, by definition, $s\sgn(0)=-1$. In that case, we find the differential asymmetry
\begin{equation}
    A(s,m,k_3,k_\perp)=-s\frac{2(v_3-n_3)\kappa(v)}{(v_3-n_3)^2+\kappa^2(v)}\approx \frac{s}{K}\frac{1-2K^2\de n_k}{1+2K^2\de n_k^2}\big[1+K^2\de n_k^2\big]^{1/2},
\end{equation}
the projection of the total angular momentum
\begin{equation}\label{projection_am_onax}
    dJ_3(s,k_3,k_\perp)=\frac{e^2}{16\pi^2}\frac{v_3-n_3}{\kappa(v)}\frac{-4sq^2}{(1-q^2)^2}\frac{dk_3 dk_\perp}{k_0 k_\perp},
\end{equation}
and the average number of twisted photons
\begin{equation}\label{av_num_onax}
    dP(s,k_3,k_\perp)=\frac{e^2}{16\pi^2}\Big[\Big(1+\frac{(v_3-n_3)^2}{\kappa^2(v)}\Big) \frac{2q^2}{1-q^2} +\Big(\frac{v_3-n_3}{\kappa(v)}+n_3 \Big)^2\Big]\frac{dk_3 dk_\perp}{k_0 k_\perp}.
\end{equation}
The last expression does not depend on the photon helicity. The second term in the square brackets in \eqref{av_num_onax} can be neglected in comparison with the first one when $q\rightarrow1$. Then the projection of the angular momentum per photon is given by
\begin{equation}\label{ell_onaxis}
    \ell(s,k_3,k_\perp)=\frac{dJ_3(s,k_3,k_\perp)}{dP(s,k_3,k_\perp)}=\frac{A}{1-q^2}\approx\frac{s}{2}\frac{1-2K^2\de n_k}{1+2K^2\de n_k^2}.
\end{equation}
The magnitude of this expression reaches its maximal value at $\de n_k\approx\pm1/(\sqrt{2}K)$, where
\begin{equation}\label{mom_per1_onax}
    \ell(s,k_3,k_\perp)\approx\mp\frac{sK}{2\sqrt{2}}.
\end{equation}
Notice that if one considers $r$ charged particles with the trajectories of the form \eqref{IR_traj} for $t>0$ with $\spx_0=0$ such that these trajectories pass to each other under the rotation by an angle of $2\pi/r$ then the property \eqref{probabil_sym} is valid. In this case, the projection of the angular momentum per photon remains the same as in \eqref{ell_onaxis} for $q\rightarrow1$ because
\begin{equation}
    \ell(s,k_3,k_\perp)=\frac{rA}{1-q^{2r}}\approx\frac{A}{1-q^{2}},\qquad r\ll m_{max}.
\end{equation}
Nonetheless, the number of radiated photons increases by the factor $r^2$ (see Fig. \ref{on_axis4_plots}).

Let us find a rough estimate for the average number of radiated twisted photons per unit energy interval for $m\neq0$, $|m|\lesssim K$. For these projections of the total angular momentum, the average number of twisted photons is peaked near $n_\perp=\be_\perp$ (see Fig. \ref{on_axis2_plots}). From \eqref{m_max} we conclude that the characteristic width of this peak is
\begin{equation}
    \de n_\perp=K\ga^{-1}\big[m^{-2}-K^{-2}\big]^{1/2}.
\end{equation}
Since $\de k_\perp=k_0\de n_\perp$, we can replace
\begin{equation}
    \frac{dk_3dk_\perp}{k_0k_\perp}\rightarrow \big[m^{-2}-K^{-2}\big]^{1/2}\frac{dk_0}{k_0}
\end{equation}
in \eqref{dP_approx}. The factor in the parenthesis in \eqref{dP_approx} is approximately equal to unity at the peak. Hence,
\begin{equation}\label{av_num_peak_onax}
    dP(s,m,k_0)\sim\frac{\al}{4\pi}e^{-2|m|K^{-1}}\big[m^{-2}-K^{-2}\big]^{1/2}\frac{dk_0}{k_0}.
\end{equation}
The average number of photons diverges in the infrared as it should be for the edge radiation.

As for the average number of twisted photons \eqref{aver_num_ph}, we see that it oscillates with the period
\begin{equation}\label{oscill_per}
    T_m=2\pi/|\de_{12}|.
\end{equation}
In particular, in case of the elastic reflection from the target located on the detector axis, when $|\mathbf{v}|=|\mathbf{u}|$, $v_3=u_3$, and $\de_{12}=\pi$, the average number of twisted photons vanishes for even $m$ (see Fig. \ref{on_axis2_plots}). The projection of the total angular momentum reads
\begin{multline}\label{dJ3_onax}
    dJ_3(s,k_3,k_\perp)=-2s\frac{e^2}{16\pi^2}\Big\{ \frac{v_3-n_3}{\kappa(v)}\frac{2q^2(v)}{(1-q^2(v))^2}+\frac{u_3-n_3}{\kappa(u)}\frac{2q^2(u)}{(1-q^2(u))^2}-\\ -\Big(\frac{v_3-n_3}{\kappa(v)} +\frac{u_3-n_3}{\kappa(u)} \Big)
    \Big(\frac{q(v)q(u)e^{i\de_{12}}}{(1-q(v)q(u)e^{i\de_{12}})^2} +c.c. \Big) \Big\}\frac{dk_3 dk_\perp}{k_0 k_\perp},
\end{multline}
and the total average number of radiated twisted photons is found to be
\begin{multline}\label{dP_onax}
    dP(s,k_3,k_\perp)=\frac{e^2}{16\pi^2}\Big[\Big(1+\frac{(v_3-n_3)^2}{\kappa^2(v)}\Big) \frac{2q^2(v)}{1-q^2(v)} +\Big(1+\frac{(u_3-n_3)^2}{\kappa^2(u)}\Big) \frac{2q^2(u)}{1-q^2(u)}-\\ -2\Big(\frac{v_3-n_3}{\kappa(v)}\frac{u_3-n_3}{\kappa(u)} +1\Big)\Big(\frac{q(u)q(v)e^{i\de_{12}}}{1-q(u)q(v)e^{i\de_{12}}} +c.c.\Big) +\Big(\frac{v_3-n_3}{\kappa(v)} -\frac{u_3-n_3}{\kappa(u)} \Big)^2 \Big]\frac{dk_3 dk_\perp}{k_0 k_\perp}.
\end{multline}
Obviously, it is independent of the photon helicity. The projection of the angular momentum per photon is given by the ratio of \eqref{dJ3_onax} and \eqref{dP_onax}.

The plots of the average number of twisted photons and the projection of the total angular momentum per photon are presented in Fig. \ref{on_axis4_plots} for several processes of scattering of electrons by the target lying on the detector axis.

\subsection{Trajectories with a break out of the detector axis}\label{Process_OffAx}

Now we investigate the production of twisted photons by charged particles moving along the trajectories \eqref{IR_traj} with the break out of the detector axis. In contrast to the case of scattering by the target lying on the detector axis, the problem at issue possesses a distinguished length scale $|x_+|$. Therefore, the average number of radiated twisted photons will depend nontrivially on the energy of a radiated photon, which, of course, will complicate the analysis.

As we have discussed above, it is sufficient to consider the radiation amplitude corresponding to the part of the trajectory with positive $t$. In a general case, the integrals entering into \eqref{probabil},
\begin{equation}\label{integrls_off}
\begin{split}
    I_3&:=\int_0^\infty dt v_3e^{-ik_0t(1-n_3v_3)+ik_3x_3}j_m\big(k_\perp (v_+t+x_+),k_\perp (v_-t+x_-)\big),\\
    I_\pm&:=\frac{in_\perp}{s\mp n_3}\int_0^\infty dt v_\pm e^{-ik_0t(1-n_3v_3)+ik_3x_3}j_{m\mp1}\big(k_\perp (v_+t+x_+),k_\perp (v_-t+x_-)\big),
\end{split}
\end{equation}
seem not to be expressible in a closed form in terms of the known special functions. Notice that the integrals of the same type arose in \cite{SIFSS} in studying the scattering of twisted electrons on the screened Coulomb potential. However, their properties were not investigated there.

\subsubsection{Exact formulas}

In the particular case when the particle accelerates instantaneously from a state of rest and then moves parallel to the detector axis, the integrals \eqref{integrls_off} are readily evaluated
\begin{equation}
    I_3+\frac12(I_++I_-)=-iv_3\frac{j_m(k_\perp x_+,k_\perp x_-)}{k_0(1-n_3v_3)}e^{ik_3x_3}.
\end{equation}
A similar expression is obtained for the amplitude of the twisted photon production in the processes of the instantaneous stopping of a charged particle (see the remark after formula \eqref{IR_traj}). This expression also describes, with good accuracy, the amplitude of radiation of twisted photons with $n_\perp\sim1/\gamma$ in the case $\be_\perp\ga=K\ll1$. The corresponding average number of twisted photons is written as
\begin{equation}\label{dP_parallel}
    dP(s,m,k_3,k_\perp)=\frac{e^2v_3^2n_\perp^3}{(1-n_3v_3)^2}|j_m(k_\perp x_+,k_\perp x_-)|^2\frac{dk_3dk_\perp}{16\pi^2 k_0^2}=\frac{e^2v_3^2n_\perp^3}{(1-n_3v_3)^2}J_m^2(k_\perp |x_+|)\frac{dk_3dk_\perp}{16\pi^2k_0^2}.
\end{equation}
We see that it is proportional to the modulus squared of the third component of the twisted photon ``wave function''. It is independent of the photon helicity, symmetric with respect to $m\rightarrow-m$, and possesses a maximum at (see, e.g., \cite{NIST})
\begin{equation}
    k_\perp|x_+|\approx |m|+0.81|m|^{1/3}+\cdots,\qquad |m|\geq1,
\end{equation}
where
\begin{equation}
    J_m^2(k_\perp |x_+|)\approx0.46|m|^{-2/3}.
\end{equation}
For $|m|\gtrsim k_\perp|x_+|$, the function $J_m^2(k_\perp |x_+|)$ goes exponentially fast to zero.

\begin{figure}[!t]
\centering
\includegraphics*[align=c,width=0.36\linewidth]{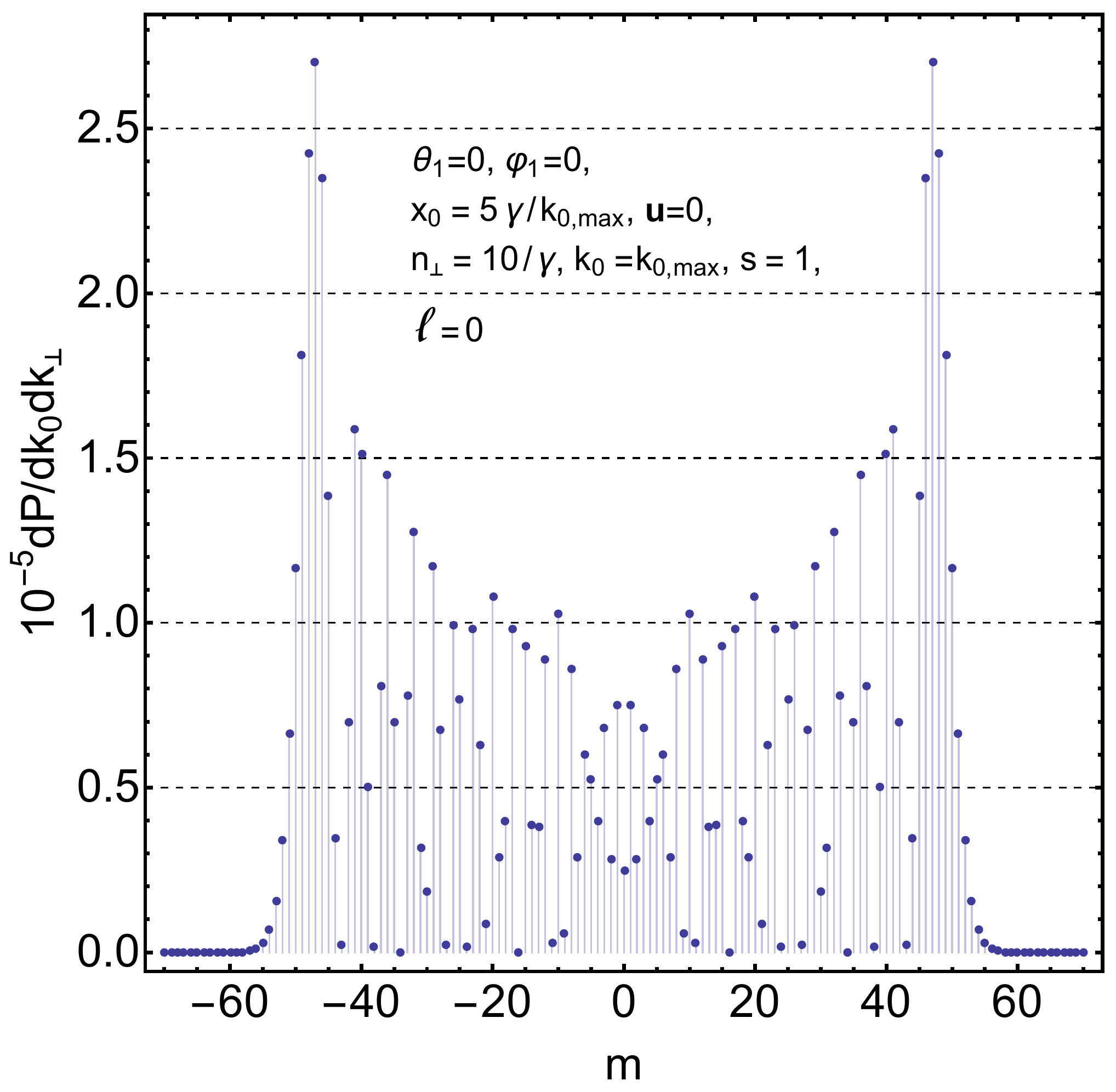}\qquad\quad
\includegraphics*[align=c,width=0.36\linewidth]{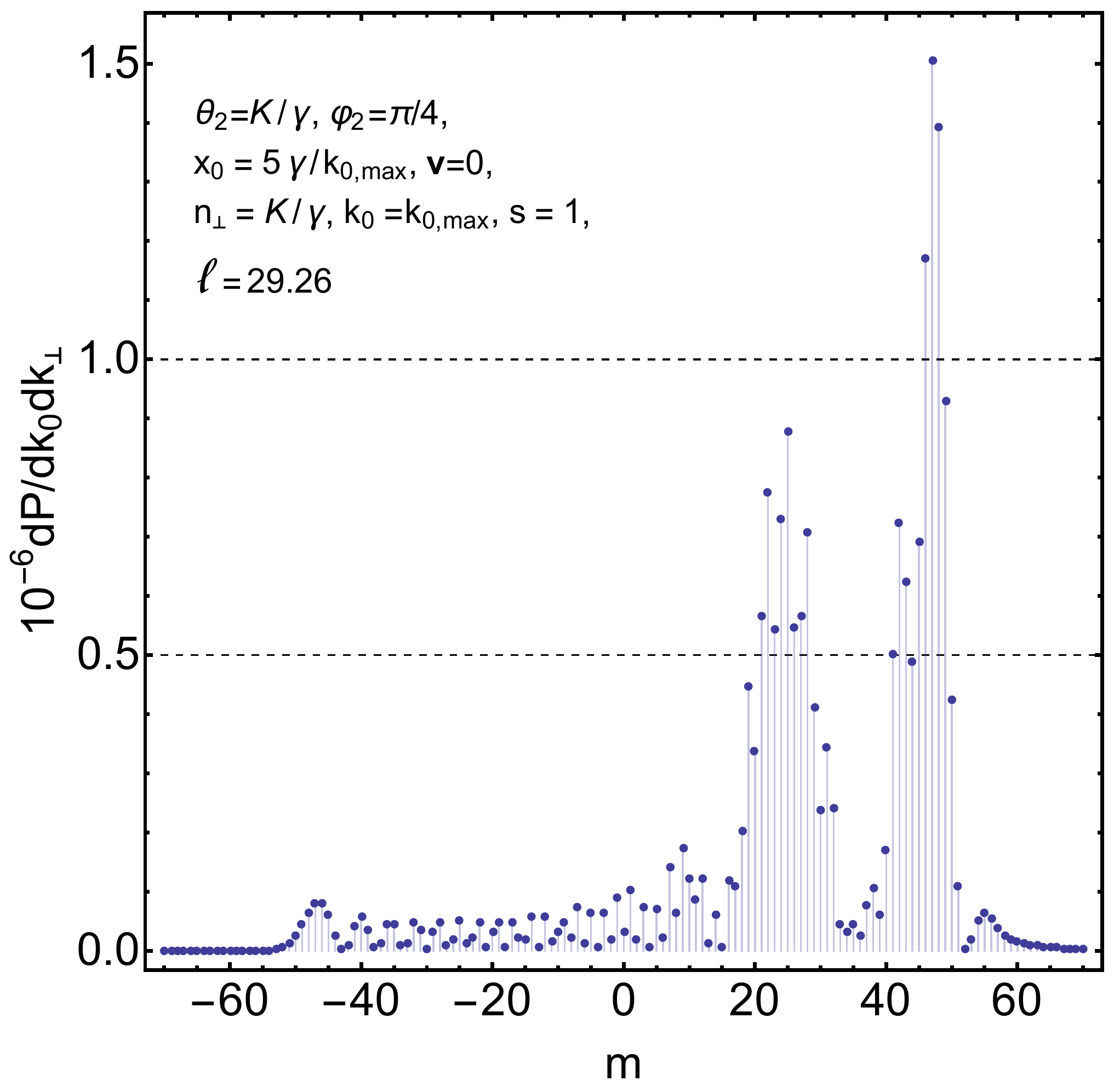}\\
\caption{{\footnotesize The average number of twisted photons against the projection of the total angular momentum produced in the elastic scattering of the electron off the target positioned at $(x_0,0,0)$, where $x_0=5\ga/k_{0,max}\approx 11$ $\mu$m. The parameters $\ga$, $K$, and $k_{0,max}$ and the definitions of the angles $\theta_{1,2}$, $\vf_{1,2}$ are the same as in Fig. \ref{on_axis2_plots}. The parameters $k_\perp|x_+|=50$ and $m_0\approx 35$. The distributions over $m$ satisfy the bound \eqref{m_int_2}. On the left panel: The production of twisted photons in the process of instantaneous stopping, the charged particle moving parallel to the detector axis. The distribution over $m$ is described by \eqref{dP_parallel}. On the right panel: The production of twisted photons in the process of instantaneous acceleration. The local minimum near $m=m_0$ and the behaviour of $dP(m)$ in the vicinity of this point are well described by \eqref{dP_offaxis_2}. The average projection of the angular momentum per photon coincides with \eqref{ell_offax}, \eqref{ell_offax_ap}.}}
\label{off_axis2_plots}
\end{figure}

Now we turn to the general case. Employing the integral representation
\begin{equation}\label{Bessel_int}
    j_m(p,q)=\int_{|z|=1}\frac{dz}{2\pi i}z^{-m-1}e^{\frac12(pz-\frac{q}{z})},
\end{equation}
we find
\begin{equation}\label{I_3_off}
    I_3=2\frac{i^{m-1}v_3}{k_\perp\be_\perp}e^{im\de+ik_3x_3}\int_{|z|=1}\frac{dz}{2\pi i}\frac{z^{-m}}{(z+q)(z+q^{-1})}\exp\Big[k_\perp x_\perp\frac{z-z^{-1}}{2} -ik_\perp x_\parallel\frac{z+z^{-1}}{2} \Big],
\end{equation}
where
\begin{equation}
    x_\perp=|x_+|\sin(\psi-\de),\qquad x_\parallel=|x_+|\cos(\psi-\de),\qquad\psi:=\arg x_+.
\end{equation}
The quantity $|x_\perp|$ is the shortest distance from the detector axis to the trajectory of a charged particle and $x_\parallel$ is the component of the vector $x_+$ parallel to the vector $v_+$. Since
\begin{equation}
    \frac{z}{(z+q)(z+q^{-1})}=\frac{1}{q^{-1}-q}\sum_{n=-\infty}^\infty(-q)^{|n|}z^{-n},
\end{equation}
we deduce with the aid of \eqref{Bessel_int} that
\begin{equation}
    I_3=i^{m-1}\frac{e^{im\de+ik_3x_3}}{k_\perp\be_\perp}\frac{2v_3}{q^{-1}-q} \sum_{n=-\infty}^\infty (-q)^{|n|}j_{m+n}(\zeta^*,\zeta),
\end{equation}
where $\zeta:=k_\perp (x_\perp+i x_\parallel)$. As for the integrals $I_\pm$, we have
\begin{equation}
    I_\pm=i^{m-1}\frac{e^{im\de+ik_3x_3}}{k_\perp\be_\perp}\frac{2}{q^{-1}-q}\frac{n_3\pm s}{n_k}\sum_{n=-\infty}^\infty (-q)^{|n\pm1|}j_{m+n}(\zeta^*,\zeta).
\end{equation}
Thus, we obtain
\begin{equation}\label{ampl_offax}
    I_3+\frac12(I_++I_-)=i^{m-1}\frac{e^{im\de+ik_3x_3}}{k_\perp\be_\perp}\frac{2q}{1-q^2}\sum_{n=-\infty}^\infty d_n j_{m+n}(\zeta^*,\zeta),
\end{equation}
where
\begin{equation}
    d_n:=(-1)^n\big( v_3q^{|n|}-\frac{n_3+s}{2n_k}q^{|n+1|}-\frac{n_3-s}{2n_k}q^{|n-1|}\big).
\end{equation}
The average number of twisted photons corresponding to the amplitude \eqref{ampl_offax} becomes
\begin{equation}\label{av_num_offax}
    dP(s,m,k_3,k_\perp)=\frac{e^2q^2}{\be_\perp^2(1-q^2)^2} \sum_{n,n'=-\infty}^\infty d_n d_{n'}j_{m+n}(\zeta^*,\zeta)j_{m+n'}(\zeta,\zeta^*)  n_\perp\frac{dk_3 dk_\perp}{4\pi^2 k_0^2}.
\end{equation}
Inasmuch as the Bessel functions tend rapidly to zero when the magnitude of their index is larger than the magnitude of their argument, the sums over $n$, $n'$ run effectively over the domain
\begin{equation}
    |m+n|\lesssim k_\perp|x_+|,\qquad |m+n'|\lesssim k_\perp|x_+|.
\end{equation}
Moreover, from \eqref{q_app}, \eqref{m_max} we see that
\begin{equation}
    |n|\lesssim m_{max},\qquad |n'|\lesssim m_{max}.
\end{equation}
These estimates can be employed, in particular, for the fast numerical evaluation of \eqref{av_num_offax}.

Formula \eqref{av_num_offax} allows us to obtain the exact expression for the average projection of the total angular momentum of radiation
\begin{equation}
    dJ_3(s,k_3,k_\perp)=\sum_{m=-\infty}^\infty mdP(s,m,k_3,k_\perp).
\end{equation}
Using the addition theorem for the functions $j_m(p,q)$ [Eq. (208), \cite{BKL2}], the recursion relations [Eq. (205), \cite{BKL2}], and the property
\begin{equation}
    j_m(0,0)=\de_{m,0},
\end{equation}
we find after a little algebra that
\begin{equation}\label{dJ3}
\begin{split}
    dJ_3(s,k_3,k_\perp)=&\,\frac{e^2q^3}{(1-q^2)^3} \Big\{-k_\perp x_\perp \Big[(n_3-n_k v_3)^2 +\frac{n_\perp^2}{4}(1-q^2)^2 -n_3n_kv_3 (1-q)^2 \Big(\frac{\kappa(v)}{n_\perp\be_\perp}-1 \Big) \Big]+\\
    &+\frac{s}{2}\frac{1-q^2}{q}\frac{n_3-v_3}{\kappa(v)}\Big\} n_\perp\frac{dk_3 dk_\perp}{2\pi^2 k_0^2}.
\end{split}
\end{equation}
The last term in this expression proportional to $s$ is exactly \eqref{projection_am_onax} as it should be. The average number of twisted photons $dP(s,k_3,k_\perp)$ coincides with \eqref{av_num_onax}. The projection of the angular momentum per photon reads
\begin{equation}\label{ell_offax}
    \ell(s,k_3,k_\perp)=\frac{dJ_3(s,k_3,k_\perp)}{dP(s,k_3,k_\perp)}.
\end{equation}
In the leading orders in $\gamma^{-1}$ and $K^{-1}$, it is given by
\begin{equation}\label{ell_offax_ap}
    \ell(s,k_3,k_\perp)\approx -k_\perp x_\perp\Big[1-\frac{2}{K}\frac{(1+K^2\de n_k^2)^{3/2}}{1+2K^2\de n_k^2}\Big] +\frac{s}{2}\frac{1-2K^2\de n_k}{1+2K^2\de n_k^2},
\end{equation}
where we have retained the next to leading in $K^{-1}$ contribution in the term at $k_\perp x_\perp$. On summing over the helicities of radiated photons in $dJ_3$ and $dP$, the term at $s$ in \eqref{ell_offax_ap} just drops out. The magnitude of $\ell$ reaches its maximum value at $|\de n_k|=1/(\sqrt{2}K)$, where
\begin{equation}\label{ell_max}
    |\ell_{max}|\approx k_\perp|x_\perp|\Big[1-K^{-1}\Big(\frac32\Big)^{3/2}\Big]+\frac{K}{2\sqrt{2}}.
\end{equation}

Formula \eqref{av_num_offax} is not very useful for an analysis of the dependence $dP(m)$ in the case when the number of relevant terms in the series in \eqref{av_num_offax} is large. Therefore, we shall find the approximate expressions for the average number of twisted photons \eqref{av_num_offax} in two instances:
\renewcommand{\theenumi}{\roman{enumi}}
\renewcommand{\labelenumi}{\theenumi)}
\begin{enumerate}
  \item $k_\perp|x_+|$ is small in comparison with $[\de n_k^2+K^{-2}]^{-1/2}$;
  \item $k_\perp|x_+|$ is much larger than $[\de n_k^2+K^{-2}]^{-1/2}$.
\end{enumerate}
The more precise formulation of these conditions will be given below.

\subsubsection{Case (i)}

We start with the first case. In the integral \eqref{I_3_off}, we make a change of the variable
\begin{equation}
    z=\frac{w-ib}{w+ib},\qquad b=\frac{1+q}{1-q}\approx 2[\de n_k^2+K^{-2}]^{-1/2}.
\end{equation}
Then the integral in \eqref{I_3_off} becomes
\begin{equation}\label{I_3_off1}
    I_3=2\frac{i^{-1-m}}{k_\perp\be_\perp} \frac{e^{im\de+ik_3x_3}v_3}{\pi(q^{-1}-q)}\int_{-\infty}^\infty \frac{dw}{w^2+1}e^{S(w)},
\end{equation}
where
\begin{equation}\label{exp_notexpnd}
    S(w)=-m\ln\frac{b+iw}{b-iw}-2ik_\perp x_\perp\frac{ bw}{w^2+b^2} +ik_\perp x_\parallel\frac{b^2-w^2}{b^2+w^2}.
\end{equation}
Keeping in mind that $b$ is large, we expand the expression in the exponent up to the leading order in $b^{-1}$:
\begin{equation}\label{exp_expnd}
    S(w)\approx ik_\perp x_\parallel -2i\frac{w}{b}(m+k_\perp x_\perp).
\end{equation}
Such an approximation for the exponent is valid for
\begin{equation}\label{estimates1}
    k_\perp|x_\perp|\frac{|w|^3}{b^3}\ll1,\qquad k_\perp|x_\parallel|\frac{|w|^2}{b^2}\ll1,
\end{equation}
and
\begin{equation}\label{estimates11}
    |m|\frac{|w|^3}{b^3}\ll1.
\end{equation}
After that, the integral \eqref{I_3_off1} is easily performed by residues. Let us denote
\begin{equation}
    m_0:=-k_\perp x_\perp.
\end{equation}
In the case $m>m_0$, the integration contour should be closed in the lower half-plane, while for $m<m_0$, it should be closed in the upper half-plane. The contribution of the residue must be larger than the contribution of the part of the integral where the estimates \eqref{estimates1} do not hold, and which was, in fact, neglected. Therefore, in addition to \eqref{estimates1}, \eqref{estimates11} we have the requirement
\begin{equation}\label{estimates2}
    \exp\Big[-\frac{2}{b}|m-m_0|\Big]\gg |w|^{-2},
\end{equation}
where $|w|\gtrsim3$. This estimate follows from the fact that the modulus of the integrand in \eqref{I_3_off1} equals $|w^2+1|^{-1}$.

Having obtained the conditions \eqref{estimates1}, \eqref{estimates11}, \eqref{estimates2}, it is convenient not to expand the exponent expression in a Taylor series as in \eqref{exp_expnd} but to deform accordingly the integration contour and take into account the contribution of the residue in the initial integral \eqref{I_3_off}. Then, if the estimates \eqref{estimates1}, \eqref{estimates2} are satisfied, the condition on $|m|$ in \eqref{estimates11} can be omitted. Indeed, one can deform the integration contour in \eqref{I_3_off1} in accordance with the steepest descent of the exact exponent expression \eqref{exp_notexpnd}. This contour is determined by the location of the saddle and singular points of \eqref{exp_expnd}. Those points can be easily found and are located out of the strip $|\im w|\lesssim b$, when the conditions \eqref{estimates1} are met but the condition \eqref{estimates11} is violated. Hence, in this case, in deforming the initial integration contour to the optimal one, it crosses one of the poles of $(w^2+1)^{-1}$. The contribution of this pole gives the leading contribution to the integral \eqref{I_3_off1}. The direction of deformation of the initial integration contour is specified by the sign of $m-m_0$.

Thus, introducing the notation,
\begin{equation}
    \epsilon:=\sgn(m-m_0),
\end{equation}
we see that for $\epsilon>0$ the integration contour in \eqref{I_3_off} ought to be deformed to the domain $|z|>q^{-1}$, while in the case $\epsilon<0$ it ought to be moved to the region $|z|<q$. As a result, we have
\begin{equation}\label{int_m2}
    I_3\approx\frac{i^{-1-m}}{k_\perp\be_\perp}e^{im\de+ik_3x_3+i\eta}\frac{2v_3}{q^{-1}-q}q^{\epsilon m}e^{\epsilon\s},
\end{equation}
where
\begin{equation}\label{eta_sigma}
\begin{split}
    \eta:=&\,k_\perp x_\parallel\frac{q+q^{-1}}{2}=k_0\frac{x_\parallel}{\be_\perp}(1-n_3v_3)\approx k_0 x_\parallel\frac{1+K^2(1+n_k^2)}{2\ga K},\\
    \s:=&\,k_\perp x_\perp\frac{q-q^{-1}}{2}= -k_0\frac{x_\perp}{\be_\perp}\kappa(v)\approx -k_0 x_\perp\frac{\big[K^4(n_k^2-1)^2+2K^2(n_k^2+1)+1\big]^{1/2}}{2\ga K}\approx\\
    \approx& -k_0 x_\perp\frac{[1+K^2\de n_k^2]^{1/2}}{\ga}.
\end{split}
\end{equation}
The integrals $I_\pm$ are expressed through $I_3$. Consequently, we obtain
\begin{equation}\label{int_m1}
    I_3+\frac12(I_++I_-)\approx \frac{i^{-1-m}}{k_\perp\be_\perp}e^{im\de+ik_3x_3+i\eta}\frac{2q^{\epsilon m}e^{\epsilon\s}}{q^{-1}-q} \Big[v_3-\frac{n_3}{n_k}\frac{q^{-1}+q}{2}-\frac{\epsilon s}{n_k}\frac{q^{-1}-q}{2} \Big] ,
\end{equation}
for $|m-m_0|>1$. Using the relations
\begin{equation}
    v_3 -\frac{n_3}{n_k}\frac{q^{-1}+q}{2}=\frac{v_3-n_3}{n_\perp^2},\qquad\frac{q^{-1}-q}{2n_k}=\frac{\kappa(v)}{n_\perp^2},
\end{equation}
we can rewrite \eqref{int_m1} as
\begin{equation}\label{int_m11}
    I_3+\frac12(I_++I_-)\approx \frac{i^{-1-m}}{k_\perp n_\perp}e^{im\de+ik_3x_3+i\eta} q^{\epsilon m}e^{\epsilon\s} \Big(\frac{v_3-n_3}{\kappa(v)} -\epsilon s\Big).
\end{equation}
As for the case $|m-m_0|<1$, it is not difficult to find that
\begin{multline}\label{int_m12}
    I_3+\frac12(I_++I_-)\approx \frac{i^{-1-m}}{k_\perp n_\perp}e^{im\de+ik_3x_3+i\eta} q^{\epsilon m}e^{\epsilon\s} \Big(\frac{v_3-n_3}{\kappa(v)} -\epsilon s\Big)+\\
    +i^{-1-m}\frac{s+\epsilon n_3}{k_\perp}\frac{e^{im\de+ik_3x_3+i\eta}}{q^{-1}-q} (q^{m-\epsilon} e^\s -q^{\epsilon-m} e^{-\s} ).
\end{multline}
Taking the modulus squared of \eqref{int_m11} or \eqref{int_m12} and substituting it into \eqref{probabil}, we obtain the average number of twisted photons produced in the processes of instantaneous acceleration or stopping. In particular, for $|m-m_0|>1$, we have
\begin{equation}\label{dP_offaxis_helic}
    dP(s,m,k_3,k_\perp)\approx e^2q^{2\epsilon m}e^{2\epsilon \s}\Big[1+\frac{(v_3-n_3)^2}{\kappa^2(v)}-2\epsilon s \frac{v_3-n_3}{\kappa(v)} \Big] \frac{dk_3dk_\perp}{16\pi^2k_0 k_\perp}.
\end{equation}
Summing over the photon helicities, we deduce
\begin{equation}\label{dP_offaxis_nh}
    dP(m,k_3,k_\perp)\approx e^2q^{2\epsilon m}e^{2\epsilon \s}\Big(1+\frac{(v_3-n_3)^2}{\kappa^2(v)}\Big) \frac{dk_3dk_\perp}{8\pi^2k_0 k_\perp}.
\end{equation}
The differential asymmetry of \eqref{dP_offaxis_nh} is
\begin{equation}
    A(m,k_3,k_\perp)\approx \left\{
                              \begin{array}{ll}
                                \dfrac{q^{2\epsilon(m) m}-q^{-2\epsilon(m) m}}{q^{2\epsilon(m) m}+q^{-2\epsilon(m) m}}, & \hbox{for $\epsilon(m)=\epsilon(-m)$}; \\
                                \tanh(2\epsilon(m) \s), & \hbox{for $\epsilon(m)=-\epsilon(-m)$}.
                              \end{array}
                            \right.
\end{equation}
The average number of twisted photons \eqref{dP_offaxis_helic}, \eqref{dP_offaxis_nh} possesses a peak at $m=m_0$. Both the expressions \eqref{dP_offaxis_helic}, \eqref{dP_offaxis_nh} decrease by a factor of $e^2$ in comparison with their values at the peak when
\begin{equation}
    |m-m_0|=m_{max}.
\end{equation}
In general, formula \eqref{dP_offaxis_helic} and the properties of the average number of radiated photons are very similar to those following from \eqref{dP_approx} with $m$ shifted by $m_0$. The projection of the angular momentum per photon is presented in \eqref{ell_offax}, \eqref{ell_offax_ap}.

\begin{figure}[!t]
\centering
a)\;\includegraphics*[align=c,width=0.35\linewidth]{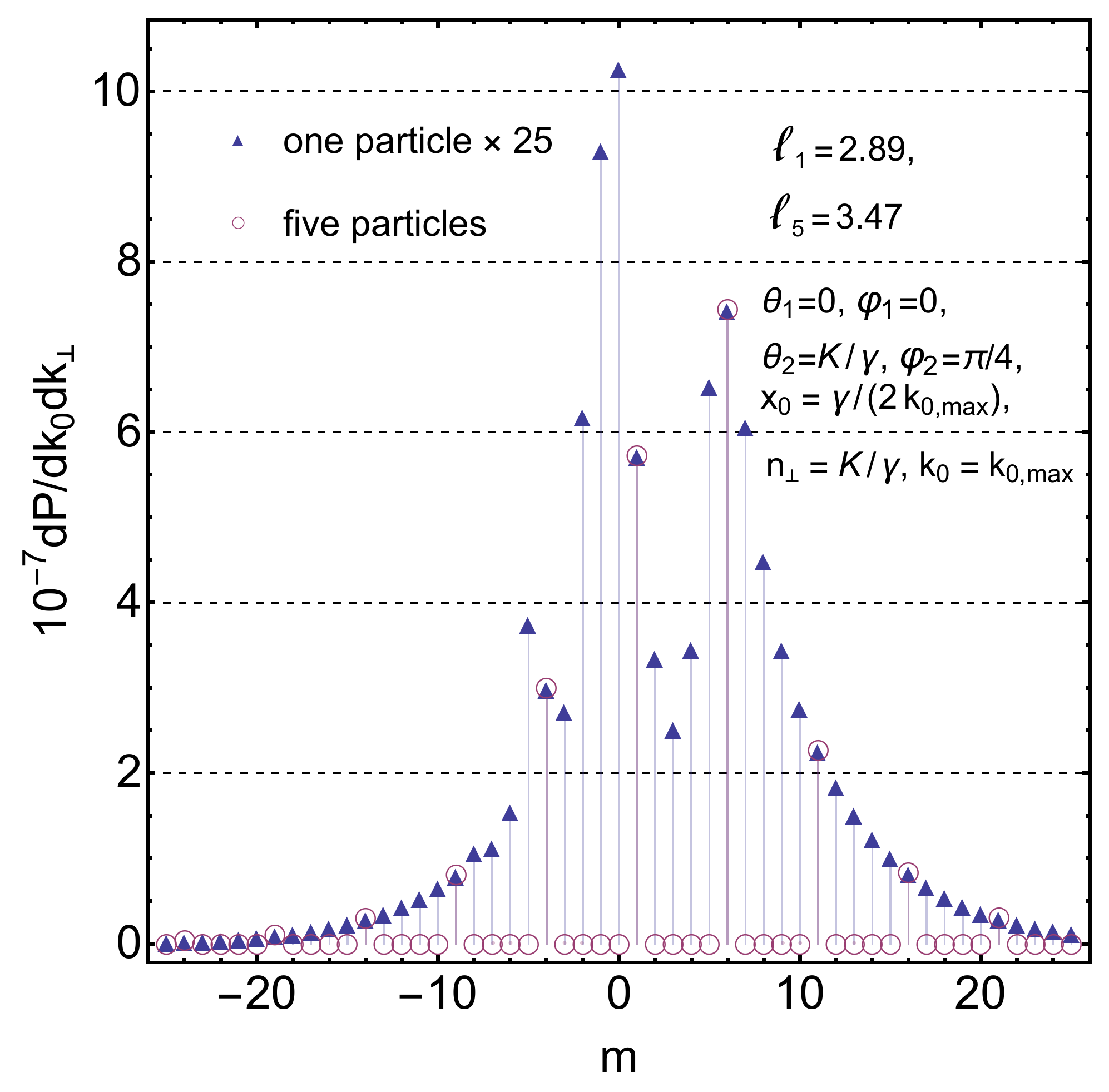}\qquad\quad
b)\;\includegraphics*[align=c,width=0.35\linewidth]{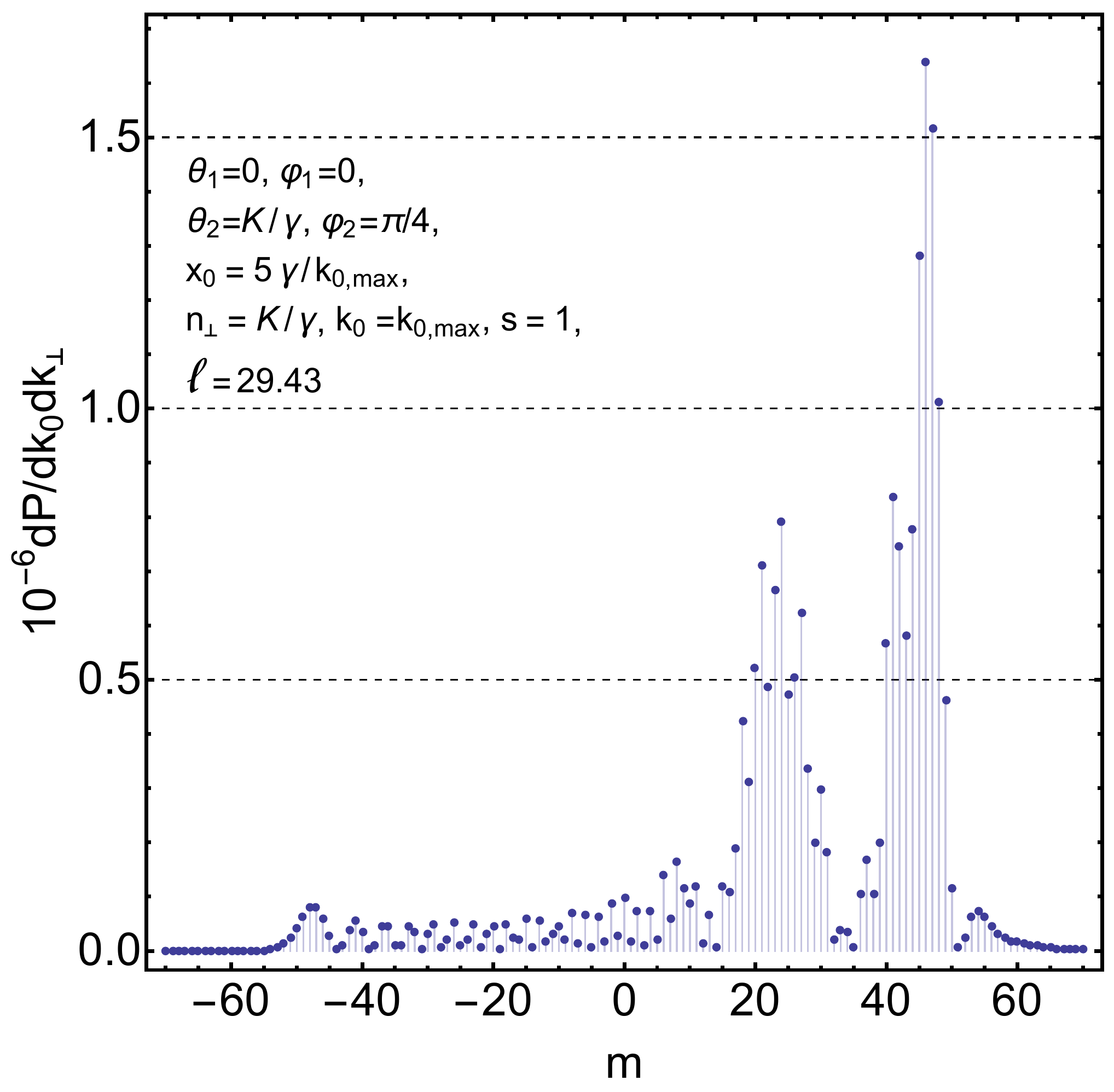}\\
c)\;\includegraphics*[align=c,width=0.35\linewidth]{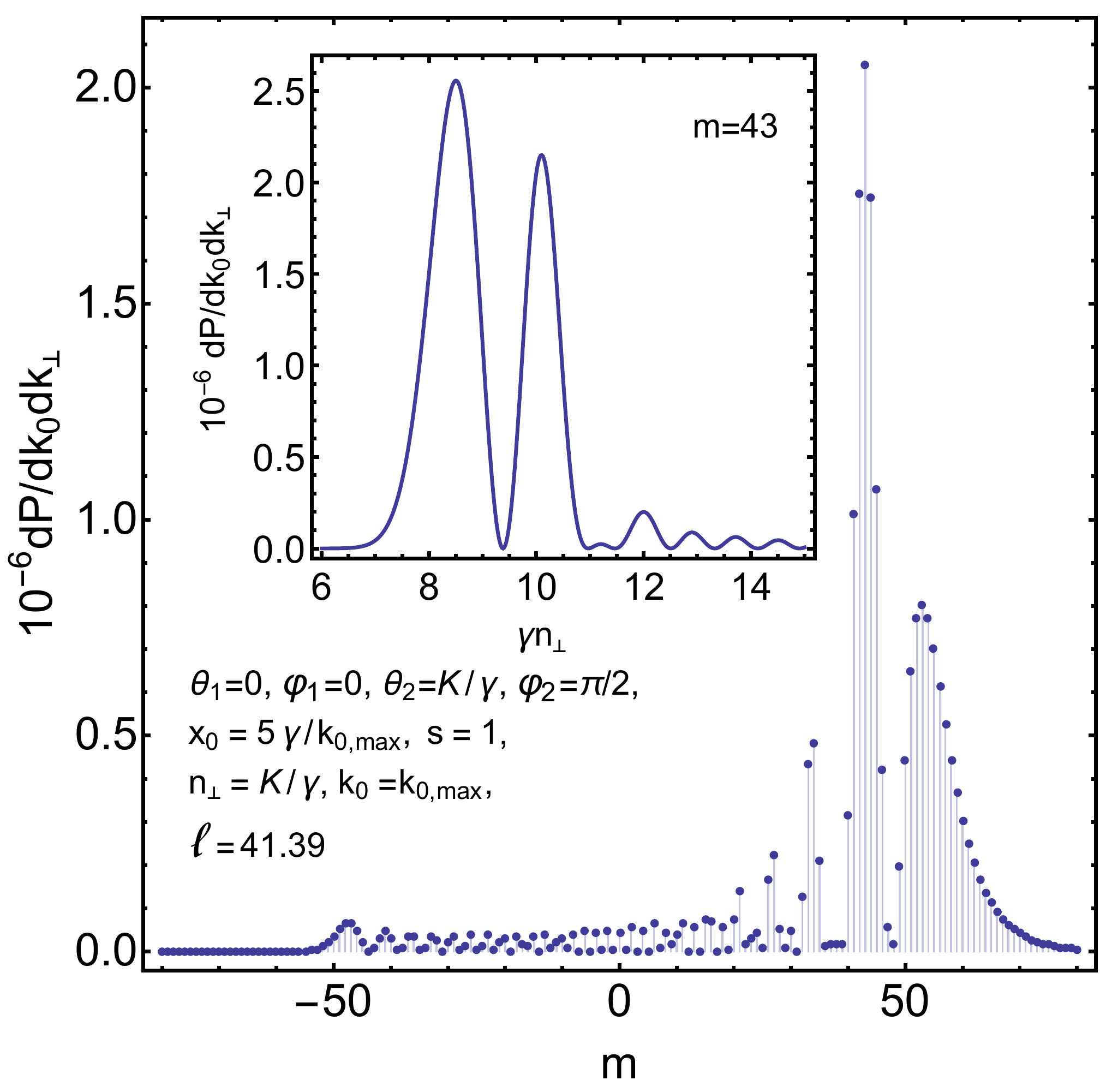}\qquad\quad
d)\;\includegraphics*[align=c,width=0.355\linewidth]{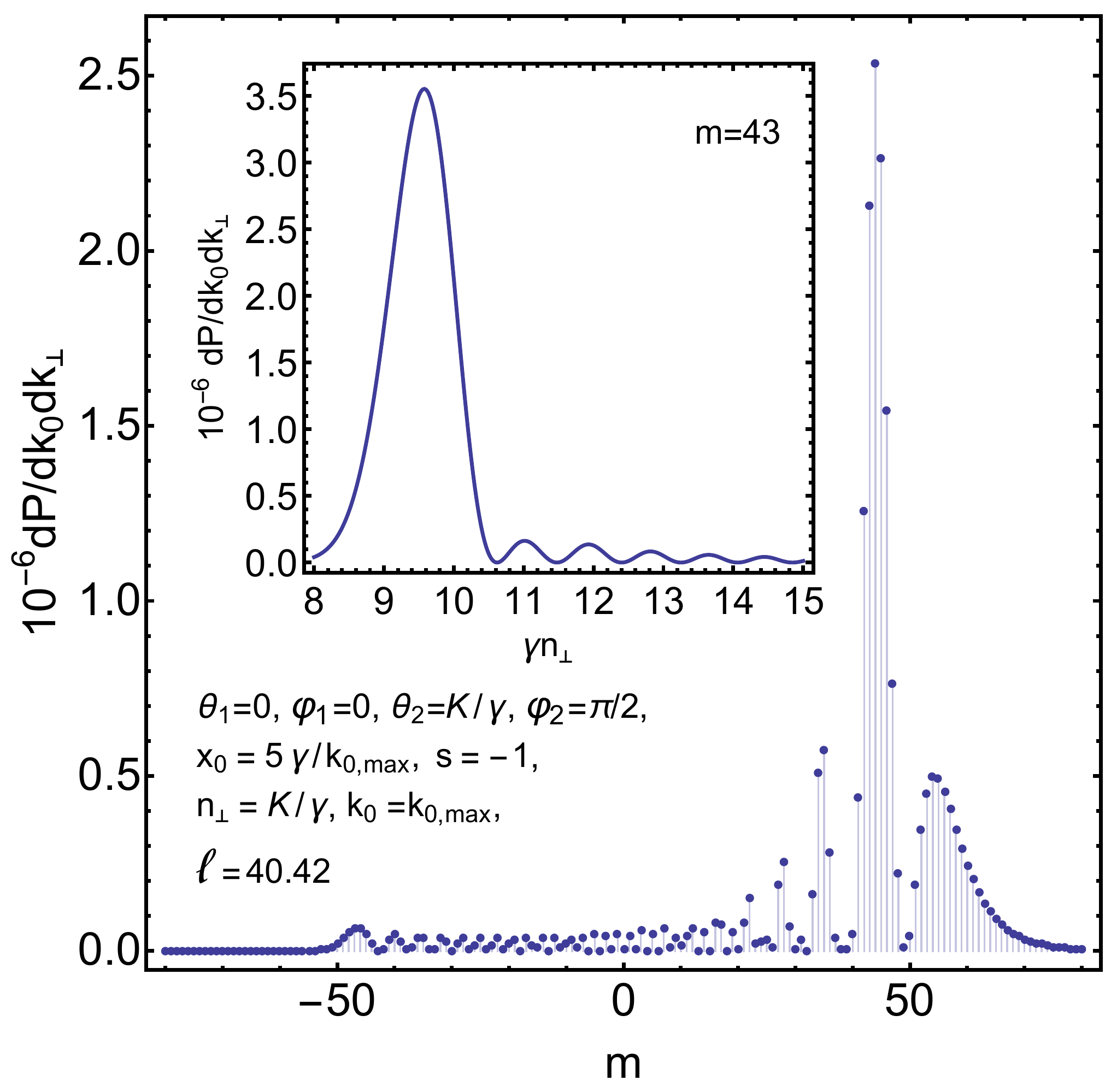}
\caption{{\footnotesize The average number of twisted photons against the projection of the total angular momentum produced in the elastic scattering of the electron off the target positioned out of the detector axis. The definitions of the parameters are the same as in Fig. \ref{off_axis2_plots}. The distributions over $m$ satisfy the bound \eqref{m_int_2}. (a) The average number of twisted photons summed over helicities. The parameters $x_0\approx1.1$ $\mu$m, $m_0\approx3.5$, and $k_\perp|x_+|=5$. The production of twisted photons in scattering of one and five electrons is compared. The trajectories of five electrons are obtained from one electron trajectory by the rotation around the detector axis, the translation along it, and the translation in time \eqref{symm_transf}. The parameter $\lambda_0\approx283$ $\mu$m is taken from \eqref{supersel_rule} such that $k_{0,max}$ is the first harmonic, and $\be_\parallel=(1-1/\gamma^2)^{1/2}$. The fulfillment of the selection rule \eqref{supersel_rule} is clearly seen. The value of the average projection of the angular momentum per photon generated by one electron is well described by \eqref{ell_offax_ap}. (b-d) The parameters $x_0\approx11$ $\mu$m and $k_\perp|x_+|=50$. The value of the average projection of the angular momentum per photon is well described by \eqref{ell_offax_ap}. The local minimum near $m=m_0\approx 35$ and the behaviour of $dP(m)$ in the vicinity of this point on the plot (b) are well described by \eqref{dP_offaxis_2}. The insets: The dependence of the average number of twisted photons on $n_\perp$ at $m=43$ and the helicities $s=\pm1$.}}
\label{off_axis4_plots}
\end{figure}

In a complete analogy with the analysis given in the previous section, we can roughly estimate the average number of twisted photons produced per unit energy scale. The result is
\begin{equation}
    dP(s,m,k_0)\sim\frac{\al}{4\pi}e^{-2|\de m|K^{-1}}\big[\de m^{-2}-K^{-2}\big]^{1/2}\frac{dk_0}{k_0},
\end{equation}
where $\de m:=m+ k_0x_\perp\be_\perp$. The plot of the average number of twisted photons for this case is given in Fig. \ref{off_axis4_plots}. The expressions \eqref{dP_offaxis_helic}, \eqref{dP_offaxis_nh} are valid when the conditions \eqref{estimates1}, \eqref{estimates2} are satisfied. However, the crude estimate for the photon energy scale where the above formulas hold can be inferred from the approximate expression \eqref{eta_sigma} for $\s$. Putting $\s\sim1$, we obtain $k_0\sim\ga/|x_\perp|$.

\begin{figure}[!t]
\centering
\includegraphics*[align=c,width=0.7\linewidth]{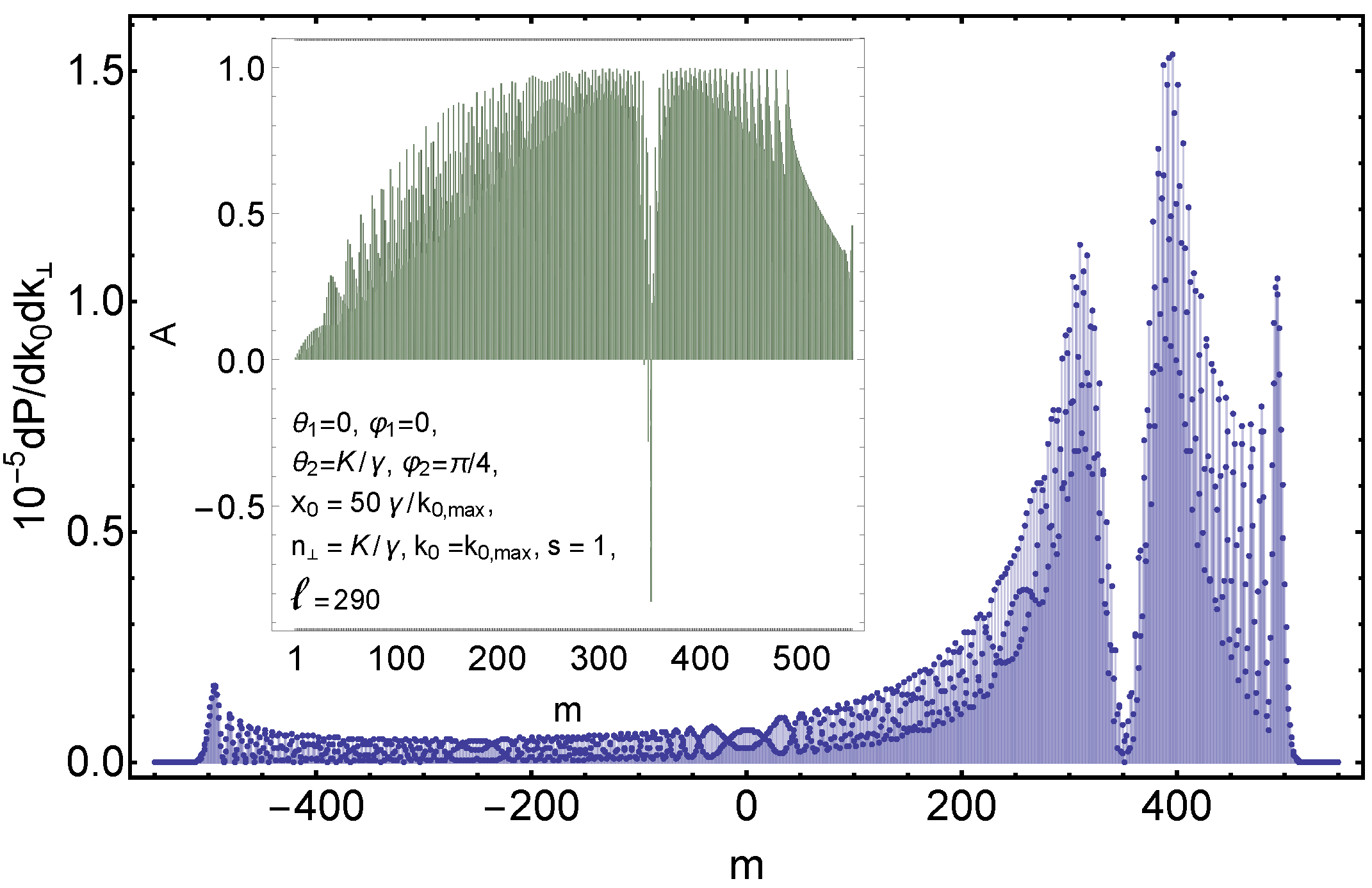}
\caption{{\footnotesize The average number of twisted photons against the projection of the total angular momentum produced in the elastic scattering of the electron off the target positioned out of the detector axis. The definitions of the parameters are the same as in Fig. \ref{off_axis2_plots}. The distribution over $m$ satisfies the bound \eqref{m_int_2}. The parameters $x_0\approx0.11$ mm, $m_0\approx354$, and $k_\perp|x_+|=500$. The local minimum near $m=m_0$ and the behaviour of $dP(m)$ in the vicinity of this point are well described by \eqref{dP_offaxis_2}. The behaviour of $dP(m)$ near $|m|=k_\perp|x_+|$ is approximately described by \eqref{dP_Ai}. The value of the average projection of the angular momentum per photon is well approximated by \eqref{ell_offax_ap}. The inset: The dependence of the differential asymmetry on $m$.}}
\label{off_axis1_plot}
\end{figure}

The second condition in \eqref{estimates1} is rather stringent. Therefore, we shall obtain below the approximate expression for the average number of radiated twisted photons when the second condition in \eqref{estimates1} is relaxed but the estimate \eqref{estimates11} is satisfied. To this end, we take into account the next term in the expansion \eqref{exp_expnd}:
\begin{equation}\label{expon_exp1}
    S(w)\approx ik_\perp x_\parallel -2i\frac{w}{b}(m+k_\perp x_\perp)-2ik_\perp x_\parallel\frac{w^2}{b^2}.
\end{equation}
Then the second condition in \eqref{estimates1} should be replaced by
\begin{equation}
    k_\perp|x_\parallel|\frac{|w|^4}{b^4}\ll1.
\end{equation}
Upon substituting \eqref{expon_exp1} into \eqref{I_3_off1}, the integral $I_3$ is reduced to the known special functions with the help of the relation
\begin{equation}
    \int_{-\infty}^\infty\frac{dx}{1+x^2}e^{-a(x-b)^2}=\frac{\pi}{2}\big[F\big(\sqrt{a}(1-ib)\big)+F\big(\sqrt{a}(1+ib)\big)\big],
\end{equation}
where the principal branch of the square root is taken and
\begin{equation}
    F(z):=e^{z^2}\erfc(z).
\end{equation}
The latter function is an entire function of $z$. Hence, we have
\begin{equation}
    I_3\approx\frac{i^{-1-m}}{k_\perp\be_\perp}\frac{v_3}{q^{-1}-q}e^{im\de+ik_3x_3+ik_\perp x_\parallel+i\frac{(k_\perp x_\perp +m)^2}{2k_\perp x_\parallel}}h(m),
\end{equation}
where
\begin{equation}
    h(m):=F\Big(\frac{\sqrt{2ik_\perp x_\parallel}}{b}+\frac{k_\perp x_\perp +m}{\sqrt{2ik_\perp x_\parallel}}\Big) +F\Big(\frac{\sqrt{2ik_\perp x_\parallel}}{b}-\frac{k_\perp x_\perp +m}{\sqrt{2ik_\perp x_\parallel}}\Big).
\end{equation}
Analogously,
\begin{equation}
    I_\pm\approx -\frac{i^{-1-m}}{k_\perp\be_\perp} \frac{n_3\pm s}{n_k} \frac{h(m\mp1)}{q^{-1}-q} e^{im\de+ik_3x_3+ik_\perp x_\parallel+i\frac{(k_\perp x_\perp +m\mp1)^2}{2k_\perp x_\parallel}}.
\end{equation}
The case of small $k_\perp |x_\parallel|$ has been already investigated above. Therefore, we suppose that
\begin{equation}
    (2k_\perp |x_\parallel|)^{1/2}\gg1.
\end{equation}
Then the function $h(m)$ is a slowly varying function of $m$ and
\begin{equation}
    h(m+1)+h(m-1)\approx 2h(m),\qquad |h(m+1)-h(m-1)|\ll|h(m)|.
\end{equation}
In this case, the average number of twisted photons produced in the processes of instantaneous acceleration or stopping reads
\begin{equation}\label{dP_offaxis_2}
    dP(s,m,k_3,k_\perp)\approx \frac{e^2q^2|h(m)|^2}{\be_\perp^2(1-q^2)^2}\Big[\Big(v_3-\frac{n_3}{n_k}\cos\frac{k_\perp x_\perp +m}{k_\perp x_\parallel} \Big)^2 +n_k^{-2}\sin^2\frac{k_\perp x_\perp +m}{k_\perp x_\parallel} \Big]n_\perp\frac{dk_3dk_\perp}{16\pi^2k_0^2}.
\end{equation}
This expression is independent of the photon helicity. As opposed to the case we have considered above, $dP$ possesses a deep local minimum at $m=m_0$. The plots of the average number of twisted photons in this case are presented in Figs. \ref{off_axis2_plots}, \ref{off_axis4_plots}, \ref{off_axis1_plot}.

\subsubsection{Case (ii)}

Now we turn to the second case. Let us represent the integral \eqref{I_3_off} in the form
\begin{equation}\label{I_3_off2}
    I_3=i^{m-1}e^{im\de+ik_3x_3}\frac{v_3}{k_0}\int_{-\pi}^\pi\frac{d\vf}{2\pi}\frac{e^{-im\vf+ik_\perp(x_\perp\sin\vf-x_\parallel\cos\vf)}}{1-n_3v_3+n_\perp\be_\perp\cos\vf}.
\end{equation}
We wish to evaluate this integral by the WKB method supposing that the exponent in the integrand is a highly oscillating function of $\vf$. More precisely, we assume that
\begin{equation}\label{estimates_WKB}
    \big[\de n_k^2+K^{-2}\big]\Big(\frac{k_\perp|x_+|}{2}\Big)^{2/3}\gg1,\qquad||m|-k_\perp|x_+||^{1/2}\Big(\frac{k_\perp|x_+|}{2}\Big)^{-1/2}\ll\pi.
\end{equation}
The first condition ensures that the fraction in the integrand is a slowly varying function of $\vf$. The second condition guarantees that the stationary points of the expression standing in the exponent are close to each other, and the integral is saturated near these stationary points. This condition is satisfied for $m\approx k_\perp|x_+|$.

It is useful to develop the expression standing in the exponent in \eqref{I_3_off2} as a Taylor series in the vicinity of the point $\vf_0:=\de-\psi\pm\pi/2$:
\begin{equation}\label{expon_expan}
    -im(\de-\psi\pm\pi/2)-i \Big[(m \mp ik_\perp|x_+|)\de\vf \pm\frac{k_\perp|x_+|}{6}\de\vf^3\Big]+\ldots,
\end{equation}
where $\de\vf:=\vf-\vf_0$. The upper sign is taken for $m>0$ and the lower sign is for $m<0$. Then we shift and stretch the integration variable
\begin{equation}
    \de\vf\rightarrow\Big(\frac{k_\perp|x_+|}{2}\Big)^{-1/3}\vf.
\end{equation}
The fraction in \eqref{I_3_off2} can be approximately written as
\begin{equation}
    \frac{1}{1-n_3v_3+n_\perp\be_\perp\cos(\vf_0+(k_\perp|x_+|/2)^{-1/3}\vf)}\approx \frac{1}{1-n_3v_3 \mp n_\perp\be_\perp\sin(\de-\psi)},
\end{equation}
provided that the first estimate in \eqref{estimates_WKB} holds. The higher order terms in \eqref{expon_expan} can be omitted and integration limits can be extended to the infinite ones if the conditions \eqref{estimates_WKB} are met. Consequently, we come to
\begin{equation}
    I_3\approx \frac{v_3}{k_0} \frac{i^{m-1}(k_\perp|x_+|/2)^{-1/3}}{1-n_3v_3 \mp n_\perp\be_\perp\sin(\de-\psi)} e^{im(\psi\mp\pi/2)+ik_3x_3} \int_{-\infty}^\infty\frac{d\vf}{2\pi} e^{-i[(\pm m-k_\perp|x_+|)(k_\perp|x_+|/2)^{-1/3}\vf+\vf^3/3]}.
\end{equation}
Since
\begin{equation}
    \Ai(x)=\int_{-\infty}^\infty\frac{d\vf}{2\pi}e^{-i(x\vf+\vf^3/3)},
\end{equation}
we derive
\begin{equation}
    I_3\approx \frac{v_3}{k_0} \frac{i^{m-1}(k_\perp|x_+|/2)^{-1/3}}{1-n_3v_3 \mp n_\perp\be_\perp\sin(\de-\psi)} e^{im(\psi\mp\pi/2)+ik_3x_3} \Ai\bigg(\frac{\pm m-k_\perp|x_+|}{(k_\perp|x_+|/2)^{1/3}}\bigg).
\end{equation}
In particular, it follows from this expression that the integral $I_3$ is exponentially suppressed in the domain
\begin{equation}
    |m|\gtrsim k_\perp|x_+|+2(k_\perp|x_+|/2)^{1/3}.
\end{equation}

The integrals $I_\pm$ are expressed through $I_3$. Hence, we obtain
\begin{multline}\label{amplitude_WKB}
    I_3+\frac12(I_++I_-)\approx \frac{i^{m-1}(k_\perp|x_+|/2)^{-1/3}e^{im(\psi\mp\pi/2)+ik_3x_3}}{k_0[1-n_3v_3 \mp n_\perp\be_\perp\sin(\de-\psi)]}  \Big[v_3f(m)  \mp\Big(\frac{s+n_3}{2n_k}f(m-1) -\frac{s-n_3}{2n_k}f(m+1)\Big)\times\\
    \times \sin(\de-\psi) \pm i\Big(\frac{s+n_3}{2n_k}f(m-1) +\frac{s-n_3}{2n_k}f(m+1)\Big) \cos(\de-\psi)\Big],
\end{multline}
where, for brevity, we have introduced the notation
\begin{equation}
    f(m):=\Ai\bigg(\frac{|m|-k_\perp|x_+|}{(k_\perp|x_+|/2)^{1/3}}\bigg).
\end{equation}
As a function of $m$, the absolute value of \eqref{amplitude_WKB}, which determines the average number of radiated twisted photons, reaches the maximum when
\begin{equation}
    |m|\approx k_\perp|x_+|-(k_\perp|x_+|/2)^{1/3}.
\end{equation}
The optimal angles providing the maximal absolute value of \eqref{amplitude_WKB} are
\begin{equation}
    \de-\psi=\pm\pi/2,
\end{equation}
where the sign is taken consistently with the sign chosen in \eqref{amplitude_WKB}. These angles correspond to the case when the scattered charged particle acquires the largest projection of the angular momentum with respect to the detector axis. If the first condition in \eqref{estimates_WKB} is satisfied, then
\begin{equation}
    |f(m-1)-f(m+1)|\ll|f(m)|,\qquad f(m-1)+f(m+1)\approx f(m),
\end{equation}
and we can simplify the amplitude \eqref{amplitude_WKB}:
\begin{equation}
    I_3+\frac12(I_++I_-)\approx \frac{i^{m-1}(k_\perp|x_+|/2)^{-1/3}e^{im(\psi\mp\pi/2)+ik_3x_3}}{k_0[1-n_3v_3 \mp n_\perp\be_\perp\sin(\de-\psi)]} f(m)  \Big[v_3 \mp\frac{n_3}{n_k}\sin(\de-\psi) \pm \frac{is}{n_k}\cos(\de-\psi) \Big].
\end{equation}
Thus the average number of twisted photons produced in the processes of instantaneous acceleration or stopping reads
\begin{equation}\label{dP_Ai}
    dP(s,m,k_3,k_\perp)\approx\frac{e^2n_\perp^3(k_\perp|x_+|/2)^{-2/3}f^2(m)}{[1-n_3v_3 \mp n_\perp\be_\perp\sin(\de-\psi)]^2}  \Big\{\Big[v_3 \mp\frac{n_3}{n_k}\sin(\de-\psi)\Big]^2 + \frac{\cos^2(\de-\psi)}{n_k^2} \Big\} \frac{dk_3dk_\perp}{16\pi^2k_0^2}.
\end{equation}
It does not depend on the photon helicity. It follows from this formula that
\begin{equation}\label{ratio1}
\begin{split}
    \frac{dP(s,m,k_3,k_\perp)}{dP(s,-m,k_3,k_\perp)}\approx&\,\frac{n_k^2-2n_k\sin(\de-\psi)+1}{[K^2(n_k^2-2n_k\sin(\de-\psi)+1)+1]^2} \frac{[K^2(n_k^2+2n_k\sin(\de-\psi)+1)+1]^2}{n_k^2+2n_k\sin(\de-\psi)+1}\approx\\
    \approx&\,\frac{n_k^2+2n_k\sin(\de-\psi)+1}{n_k^2-2n_k\sin(\de-\psi)+1}.
\end{split}
\end{equation}
The cases $n_k=1$, $\sin(\de-\psi)=\pm 1$ are not well described by the above expression. In these cases, formula \eqref{dP_Ai} implies
\begin{equation}\label{ratio11}
    \frac{dP(s,m,k_3,k_\perp)}{dP(s,-m,k_3,k_\perp)}\approx\Big(\frac{K^4}{\ga^4}\Big)^{\pm1},
\end{equation}
respectively.

\subsubsection{General estimates}\label{GenEstm}

From the above analysis we see that the large angular momenta per photon are achieved for large $|x_+|$. Although the maximum value of $dP$ declines as $|x_+|^{-2/3}$ in increasing $|x_+|$, it appears that one can gain the twisted photons with almost arbitrary large projections of the total angular momentum, which are equal to $k_\perp|x_+|$ by the order of magnitude. However, one should bear in mind that, in the wave zone, where the twisted photons are detected, $|x_+|\ll R$, where $R$ is the distance from the origin to the detector. The radius of the detector that is able to record such a twisted photon should be
\begin{equation}\label{detect_rad}
    r_d\gtrsim Rn_\perp\gg |x_+|n_\perp.
\end{equation}
Consequently, at fixed $r_d$, there is the maximum value of $|x_+|$ determined by \eqref{detect_rad} above which the number of recorded twisted photons falls exponentially fast to zero.

We also can draw the general estimates of the maximum value of the projection of the angular momentum of photons radiated by a classical current. The integral entering into formula \eqref{probabil} for the average number of twisted photons can be partitioned into the three integrals: two of them correspond to the uniform rectilinear motion of the particle before and after the interaction with the external electromagnetic field, and the rest integral describes the contribution to the radiation amplitude from the part of the particle trajectory where its acceleration does not vanish. The integration limits in the latter integral are finite, and the integrand tends exponentially fast to zero out of the interval
\begin{equation}\label{m_int_1}
    |m|\lesssim k_\perp r_m,
\end{equation}
where $r_m=\max|x_+|$ is the maximal distance from the detector axis to the points of this part of the particle trajectory. As for the parts of the particle trajectory responsible for the edge radiation, we have already seen above that the spectrum of projections of the total angular momentum extends over the interval
\begin{equation}\label{m_int_2}
    |m|\lesssim k_\perp |x_+|+\be_\perp\ga.
\end{equation}
Hence, the whole spectrum of $m$'s belongs to the union of the intervals \eqref{m_int_1}, \eqref{m_int_2}.

\begin{figure}[!t]
\centering
\includegraphics*[align=c,width=0.35\linewidth]{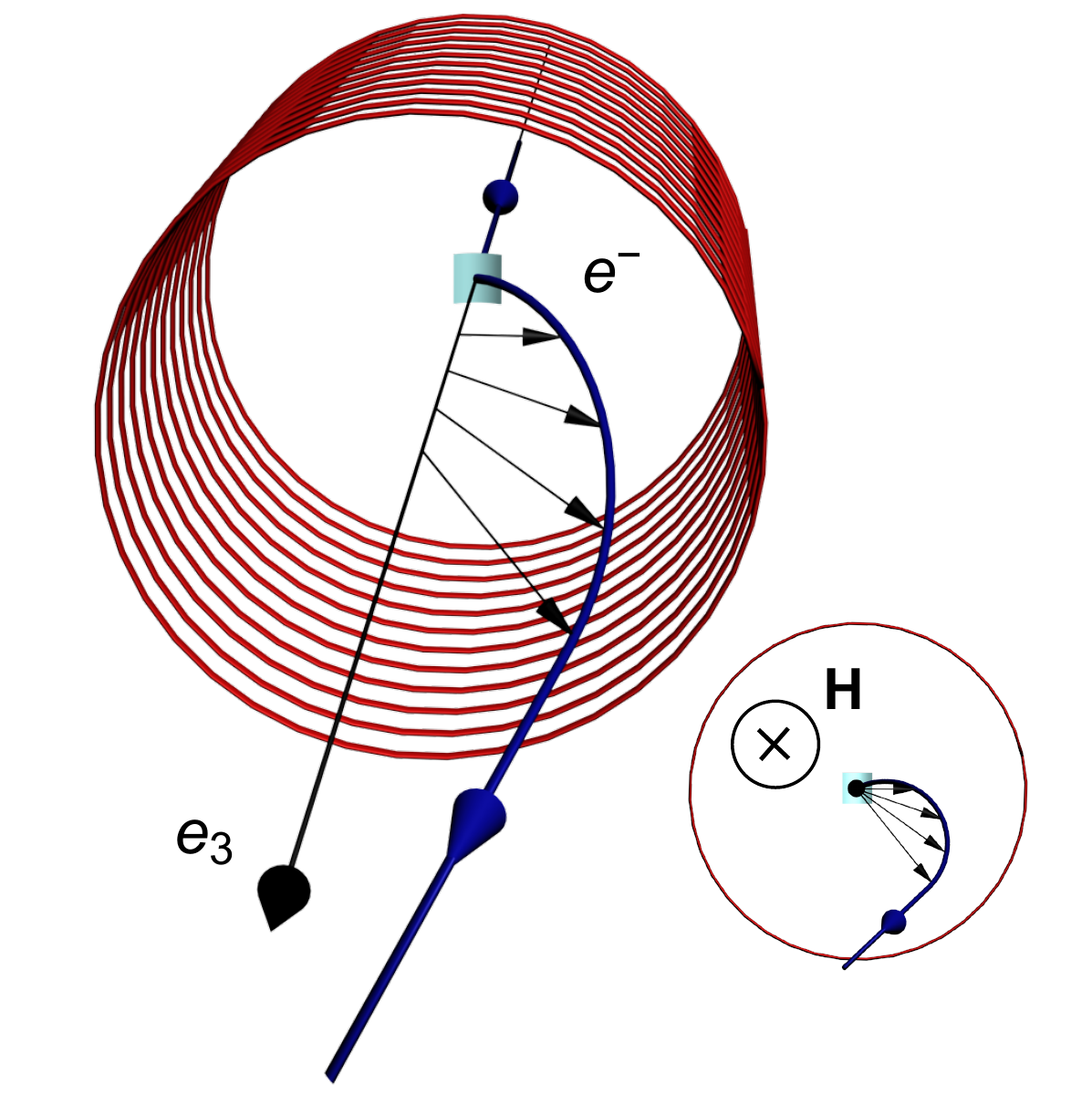}
\caption{{\footnotesize Schematic representation of the trajectory of the electron scattered on the bent crystal in the solenoid (see the detailed description in the main text). The small picture shows the direction of the magnetic field vector and the electron trajectory as they look from the detector location.}}
\label{scheme}
\end{figure}

As an example, let us consider the forward radiation of the wiggler. In this case,
\begin{equation}
    r_m=\frac{K}{\omega\ga},\qquad k_0=\frac{2n\omega\ga^2}{1+K^2(1+n_k^2)},
\end{equation}
and $|x_+|=0$, $\be_\perp=0$ in \eqref{m_int_2}. Therefore,
\begin{equation}
    k_\perp r_m=\frac{2K^2n_kn}{1+K^2(1+n_k^2)}\approx\frac{2n_k n}{1+n_k^2}.
\end{equation}
In the optimal case, $n_k=1$ and
\begin{equation}
    k_\perp r_m=n\;\;\Rightarrow\;\;|m|\lesssim n.
\end{equation}
This agrees with the results presented in \cite{BKL2}. As for the radiation at an angle, it was shown in [Eq. (101), \cite{BKL2}] that
\begin{equation}\label{undul_mmax}
    |m|\lesssim \frac{20N}{7}k_\perp\theta\omega^{-1}=\frac{10}{7\pi}k_\perp\theta\la_0 N,
\end{equation}
where $\la_0$ is the length of the undulator section. The quantity $\la_0N$ is the length of the undulator, and $\theta\la_0N/2$ defines $r_m$. We see that \eqref{undul_mmax} coincides with \eqref{m_int_1}.

\begin{figure}[!t]
\centering
a)\;\includegraphics*[align=c,width=0.7\linewidth]{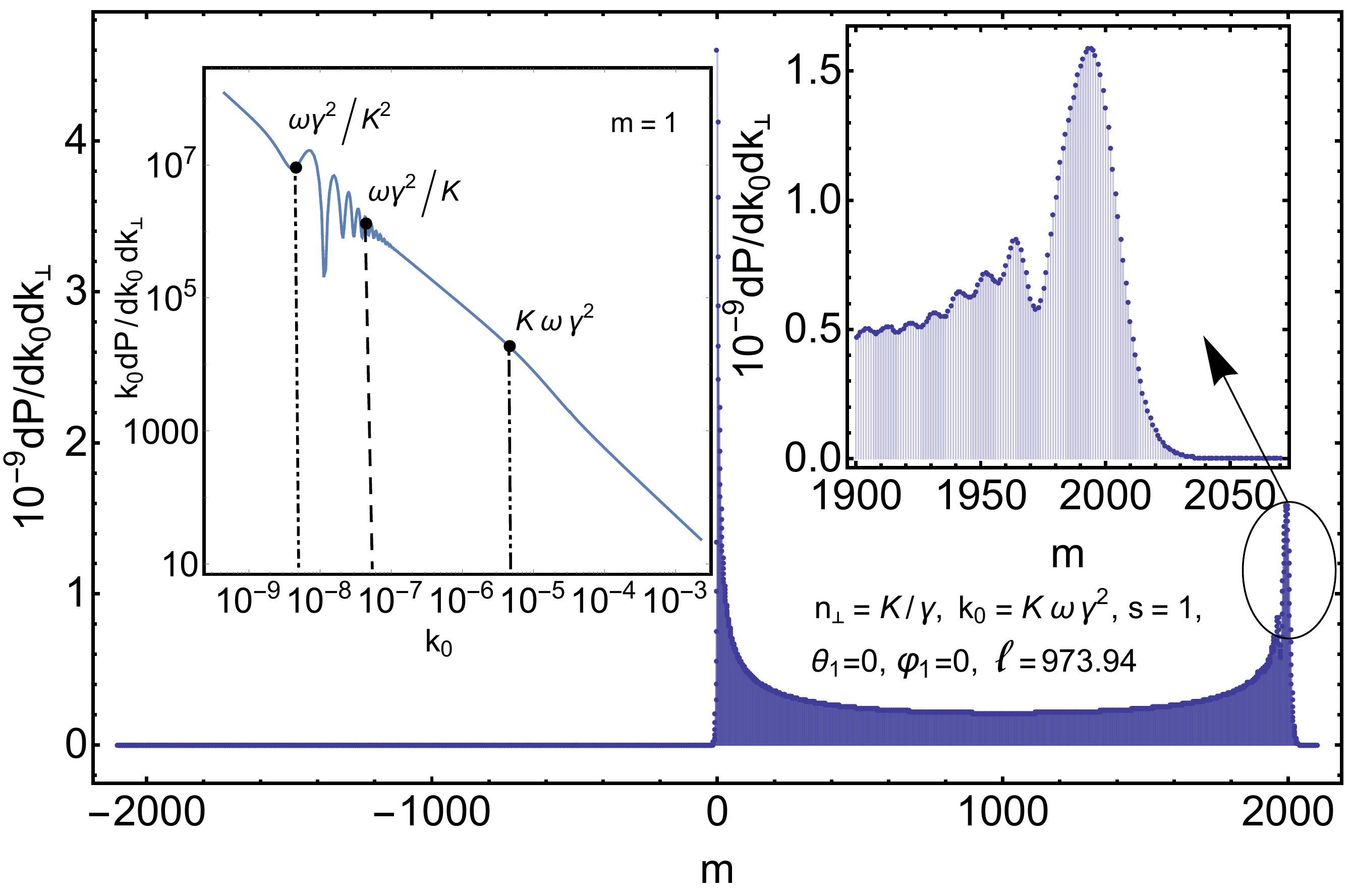}\\
b)\;\includegraphics*[align=c,width=0.35\linewidth]{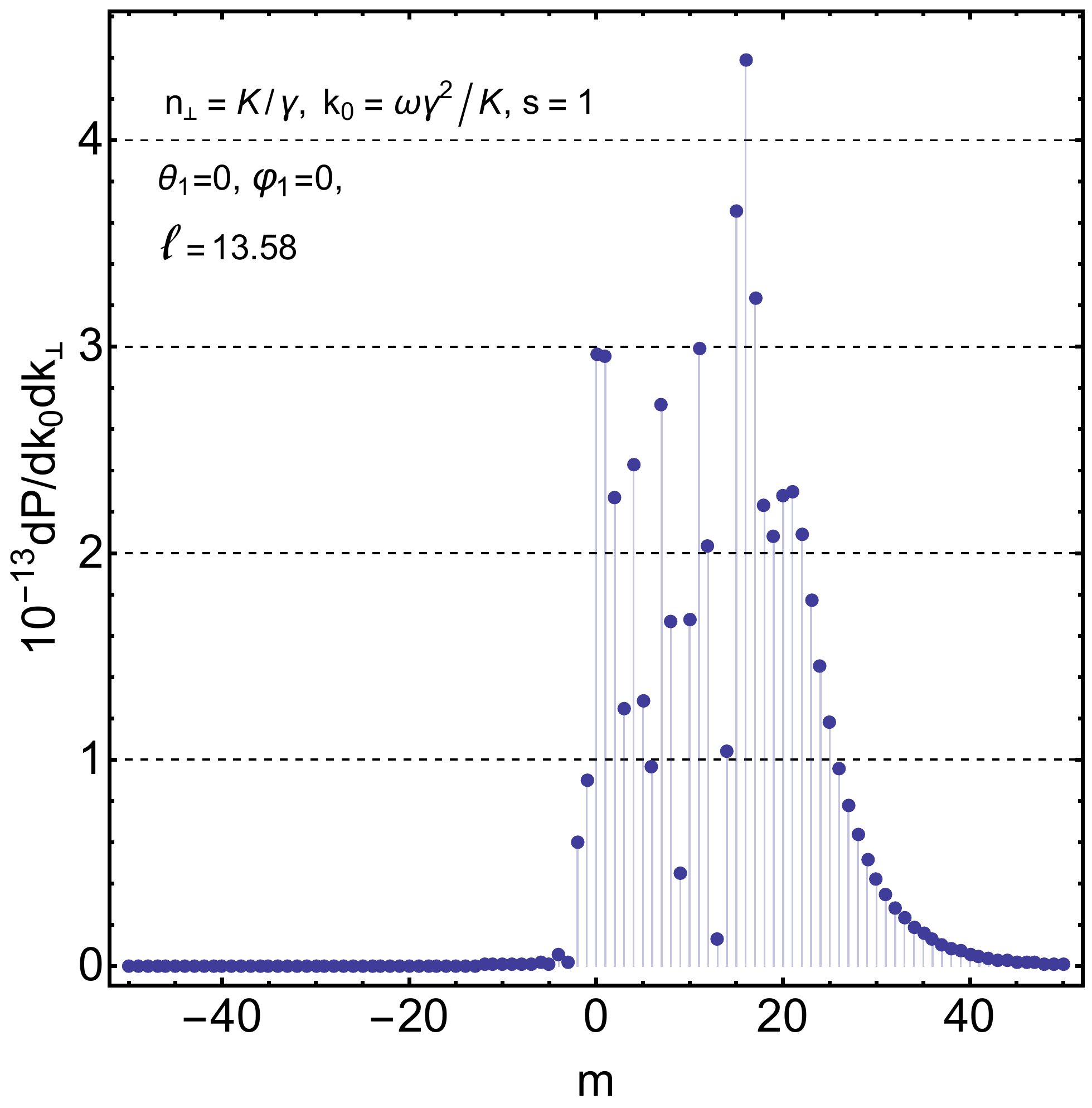}\qquad\quad
c)\;\includegraphics*[align=c,width=0.35\linewidth]{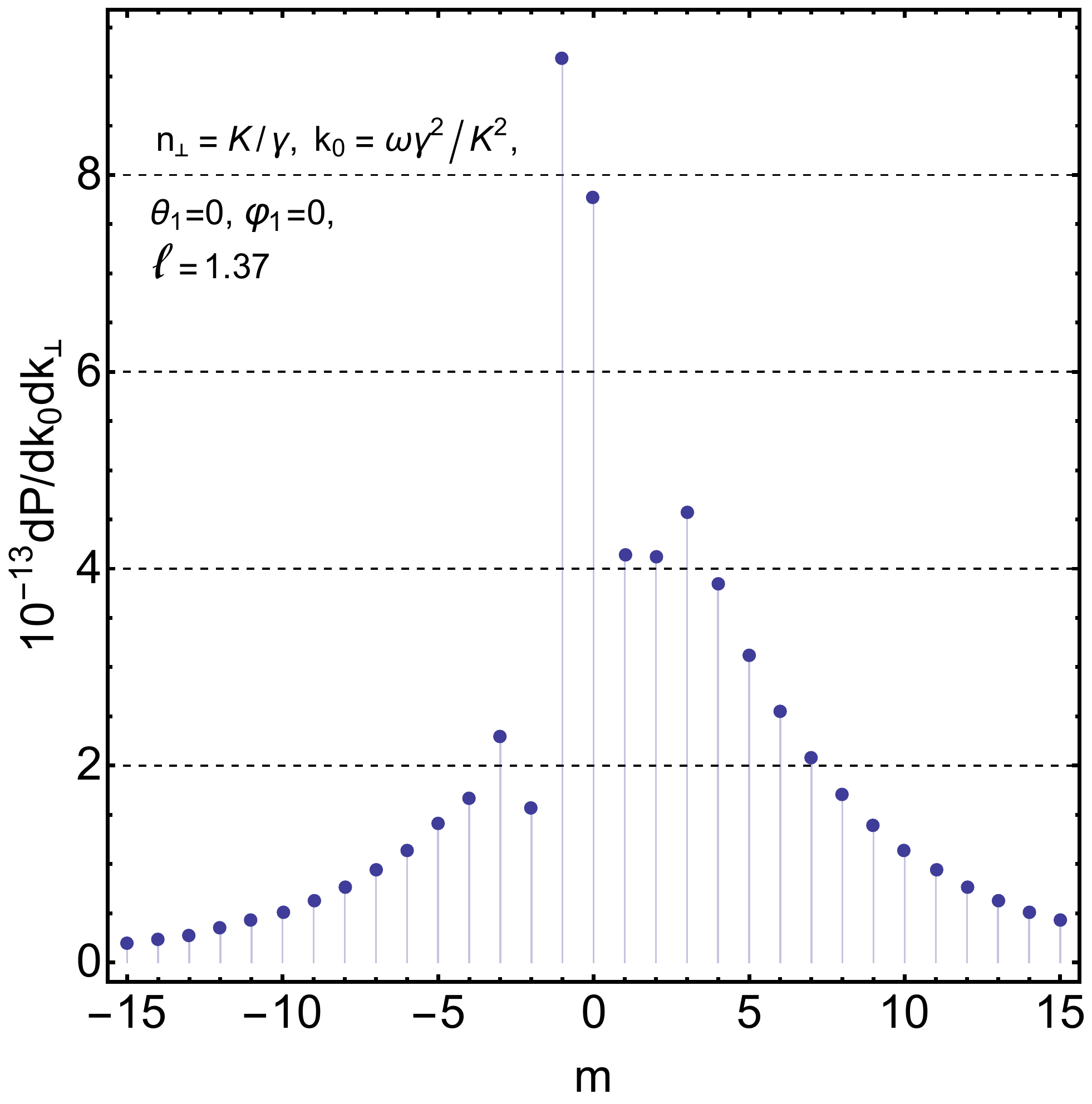}
\caption{{\footnotesize The average number of twisted photons against the projection of the total angular momentum produced in the elastic scattering of the electron off the target positioned on the detector axis in the magnetic field of a solenoid (see the detailed description in the main text). The distributions over $m$ satisfy the bound \eqref{m_int_2}. The parameters $K=10$, $\ga=10^3$, the radius of the solenoid is approximately $2\rho\approx1.7$ cm, the length of the solenoid is approximately $\pi\be_\parallel/\omega\approx 267$ cm. (a) The average number of twisted photons against the projection of the total angular momentum produced at the synchrotron scale. The parameter $2 k_\perp\rho=2000$. The inset: The dependence of the intensity of radiated twisted photons with $m=1$ on the photon energy measured in the rest energies of the electron. The three energy scales \eqref{energy_scales} are clearly seen. (b) The same as in (a) but at the near-infrared scale. The parameter $2 k_\perp\rho=20$. (c) The same as in (a) but summed over helicities and taken at the infrared scale. The parameter $2 k_\perp\rho=2$.}}
\label{magn3_plots}
\end{figure}

In Figs. \ref{off_axis2_plots}, \ref{off_axis4_plots}, \ref{off_axis1_plot}, the plots of the average number of twisted photons produced in different elastic scattering processes are presented. In Figs. \ref{off_axis2_plots}, \ref{off_axis4_plots}, \ref{off_axis1_plot}, the charged particle changes its state instantaneously at the point lying out of the detector axis. Such a trajectory models, for example, the scattering of an electron (or positron) by a crystal. As a result of this scattering, the electron acquires a nonzero angular momentum with respect to the detector axis. It equals $m x_\perp K$, where $m$ is the electron mass. In virtue of the total angular momentum conservation law, this leads to a radiation of twisted photons. Our approach describes the infrared asymptotics of this radiation. As for the radiation of hard photons in such a process, its description can be found, for example, in \cite{GuiBandTikh}. It is also possible to arrange symmetrically the scattering crystals around the detector axis such that the trajectories of scattered particles are obtained from each other either by the rotation around the detector axis by an angle of $2\pi/r$, $r\in \mathbb{N}$, or by the composition of the rotation around the detector axis, the translation along it, and the translation in time. Then the coherent radiation of twisted photons is described by \eqref{probabil_sym} or \eqref{probabil_sym1}, \eqref{interfer_factor}, \eqref{supersel_rule}, respectively. In the latter case, one can generate, in principle, much harder twisted photons than those produced in undulators since $|\la_0|$ is much smaller than the undulator section length. The incoherent radiation is given by the one-particle radiation multiplied by the number of particles in the bunch.

In Figs. \ref{scheme}, \ref{magn3_plots}, a slightly different situation is considered. The electrons move along the symmetry axis of a solenoid, which coincides with the detector axis. These electrons scatter elastically off the target located on the detector axis and acquire the perpendicular velocity component $\be_\perp\equiv K/\gamma$. Then, they move along a helix in the magnetic field produced by the solenoid (we assume that $H=2 \times 10^4$ G). On making a half-turn, the electrons escape the magnetic field and move freely along a straight line. As a result, they possess a nonzero angular momentum with respect to the detector axis, which is equal to $2K^2/H$, where $H$ is measured in the critical fields \eqref{critical_field}. This leads to the production of twisted photons. One can distinguish the four characteristic photon energy scales in this process: the synchrotron scale (see, e.g., \cite{Bord.1})
\begin{equation}
    k^{syn}_0\approx K\omega\ga^2,
\end{equation}
where $\omega=H/\ga=K/(\ga\rho)$ and $\rho$ is the radius of a circular helix; the near infrared scale $k^{nIR}_0\approx k^{syn}_0/K^2$; the infrared scale $k^{IR}_0\approx k^{syn}_0/K^3$; and the far infrared scale $k^{fIR}_0\ll k^{syn}_0/K^3$. At the synchrotron scale, the radiation is formed on a small part of the electron trajectory. At the near infrared scale, the radiation is formed on the whole trajectory and the details of the dynamics are still relevant for the properties of radiation. At the infrared and far infrared scales, we fall into the cases studied above, i.e., it is the edge radiation. These photon energy scales depend on the parameters of the installation and may not coincide with the standard classification of the different ranges of the electromagnetic spectrum. For example, for $H=2 \times 10^4$ G, $K=10$, $\ga=10^3$, we have
\begin{equation}\label{energy_scales}
    k^{syn}_0\approx 2.3\; \text{eV},\qquad k^{nIR}_0\approx 23\; \text{meV},\qquad k^{IR}_0\approx 2.3\; \text{meV},\qquad k^{fIR}_0\lesssim 0.23\; \text{meV}.
\end{equation}
Hence, we see that the photons with the energies $k^{nIR}_0$, $k^{IR}_0$ belong to the THz spectrum domain, i.e., they are in the far infrared.

\section{Conclusion}


Let us sum up the results. We studied the infrared asymptotics of the radiation of twisted photons for QED processes in a vacuum neglecting the effects of a chamber where the processes are evolving. This asymptotics is universal and can be described by the radiation of the classical current corresponding to the charged particles moving uniformly along straight lines with the break at the instant of time $t=0$. For brevity, we call the process corresponding to such trajectories as scattering. The radiation of that current is known as the edge radiation, and we, in fact, represented this radiation in terms of twisted photons. In contrast to the description of the edge radiation in terms of the plane-wave photons, the probability of radiation of twisted photons is not invariant under the translations of the origin that are perpendicular to the detector axis. Therefore, it is relevant for our analysis where the break of the trajectory is located.

We proved several general results about the radiation of twisted photons. First, we established the symmetry property \eqref{symm_prop} of the average number of twisted photons produced by the charged particles scattered off a target located on the detector axis. This symmetry property holds for any QED process in a vacuum in the far infrared. Second, we found the general expression for the upper bound \eqref{m_int_1}, \eqref{m_int_2} of the magnitude of the total angular momentum of a twisted photon radiated by a classical current. Third, we proved the selection rule \eqref{selection_rule} for the radiation (absorbtion) of twisted photons by the classical currents symmetric under the rotation by an angle of $2\pi/r$, $r\in \mathbb{N}$, around the detector axis. A similar selection rule was pointed out in \cite{RubDun08,RubDun11} for the scattering of light by microstructures. Fourth, we generalized the selection rule \eqref{selection_rule} to the case when the twisted photons are created by the classical currents of identical charged particles with trajectories obtained from each other by the rotation by an angle around the detector axis, the translation along it, and the translation in time (see \eqref{probabil_sym1}, \eqref{interfer_factor}, \eqref{supersel_rule}).

We obtained the exact formula for the probability of radiation of twisted photons produced by scattering of the charged particles off the target located on the detector axis. This allowed us to find the main characteristics of the radiation twist: the
differential asymmetry, the average projection of the total angular momentum of radiation, and the projection of the total angular momentum per photon. The maximal angular momentum per photon is given by \eqref{mom_per1_onax} in this case. It is achieved in the ultrarelativistic regime for the twisted photons with
\begin{equation}\label{n_perp_max}
    n_\perp\approx\be_\perp\pm (\sqrt{2}\ga)^{-1},
\end{equation}
where $\be_\perp$ is the perpendicular velocity component of the scattered charged particle such that $\be_\perp\ga\gtrsim3$. The probability of radiation of twisted photons appears to have a sharp peak near $n_\perp\approx\be_\perp$. We found the estimate \eqref{av_num_peak_onax} for the average number of twisted photons radiated in this peak. As a function of the total angular momentum $m$, the probability of radiation of twisted photons oscillates with the period \eqref{oscill_per}. In particular, there exists the selection rule -- $m$ is an odd number -- for the radiation of twisted photons in the process of elastic reflection of a charged particle from the detector axis. The plots of the average number of radiated twisted photons for the different scattering processes with the target lying on the detector axis are given in Figs. \ref{on_axis2_plots}, \ref{on_axis4_plots}

As for the scattering by the target lying out of the detector axis, we did not succeed in finding the exact closed formula for the average number of radiated twisted photons. Nevertheless, we represented it in the form of a series \eqref{av_num_offax} in terms of the Bessel functions, which can be used for its fast numerical evaluation. Such a representation allowed us to obtain the exact expressions for the average projection of the total angular momentum \eqref{dJ3}, and the projection of the total angular momentum per photon \eqref{ell_offax} for the processes of instantaneous stopping or acceleration of a charged particle. In that case, the maximal angular momentum per photon is given by \eqref{ell_max} and realized for the same $n_\perp$ as in \eqref{n_perp_max}. As compared with the case when the target is located on the detector axis, the angular momentum per photon acquires the addition of the order $-k_\perp x_\perp$, where $|x_\perp|$ is the shortest distance from the detector axis to the trajectory of a charged particle. We also derived the approximate expressions \eqref{dP_offaxis_helic}, \eqref{dP_offaxis_2}, and \eqref{dP_Ai} for the average number of radiated twisted photons in the different domains of quantum numbers.

Several scattering processes with the target located out of the detector axis were investigated numerically, and the respective plots of the average number of radiated twisted photons are presented in Figs. \ref{off_axis2_plots}, \ref{off_axis4_plots}, \ref{off_axis1_plot}. We also considered the production of twisted photons by electrons scattered off the target that is located on the detector axis in the magnetic field of a solenoid, the solenoid axis coinciding with the detector axis (see Fig. \ref{scheme}). For the modest electron energies and the magnetic field strengths, such a process generates the twisted photons with the energy $2.3$ eV (in the synchrotron range) and $m\approx 2000$. The projection of the total angular momentum per photon in this case $\ell\approx974$ (see Fig. \ref{magn3_plots}). The infrared asymptotics, $k_0\lesssim 2.3$ meV, of the probability of radiation of twisted photons in this process is described by the formulas we obtained.

\paragraph{Acknowledgments.}

We are thankful to V.G. Bagrov and D.V. Karlovets for the fruitful conversations. This work was supported by the Russian Science Foundation (project No. 17-72-20013).


\end{document}